\documentclass[11pt, a4paper]{article}
\usepackage{hyperref}
\usepackage{epsfig}
\usepackage[latin1]{inputenc}
\usepackage{bbm,amsfonts}
\usepackage{graphicx,tikz}
\usepackage{amssymb,amsmath,braket,comment}

\csname @addtoreset\endcsname{equation}{section}

\def\bseq{\begin{subequation}}  
\def\eseq{\end{subequation}}
\def\bsea{\begin{subeqnarray}}  
\def\esea{\end{subeqnarray}}

\newcommand{\beq}{\begin{equation}}
\newcommand{\bea}{\begin{eqnarray}}
\newcommand{\eea}{\end{eqnarray}}
\newcommand{\eeq}{\end{equation}}

\newcommand {\non}{\nonumber}

\renewcommand{\a}{\alpha}
\renewcommand{\b}{\beta}

\renewcommand{\d}{\delta}

\newcommand{\g}{\gamma}
\newcommand{\G}{\Gamma}

\newcommand{\e}{\epsilon}
\newcommand{\z}{\zeta}

\renewcommand{\l}{\lambda}

\newcommand{\m}{\mu}

\newcommand{\n}{\nu}

\newcommand{\p}{\pi}

\newcommand{\s}{\sigma}

\def\beq{\begin{equation}}
\def\eeq{\end{equation}}
\def\bea{\begin{eqnarray}}
\def\eea{\end{eqnarray}}
\def\Tr{\mathrm{Tr}}

\def\a{\alpha}
\def\b{\beta}
\def\g{\gamma}

\def\d{\delta}
\def\e{\epsilon}
\def\z{\zeta}

\def\l{\lambda}
\def\G{\Gamma}

\def\TRIANGLE{\textup{\bf T}}
\def\BOX{\textup{\bf B}}

\renewcommand{\sp}[2]{\left( {#1} \cdot {#2}\right)}

\newcommand{\sunset}{
  \begin{tikzpicture} [scale=0.5]
    \draw (0,0) circle (1) ;
    \draw (-1.5,0) -- (1.5,0);
  \end{tikzpicture}
}
\newcommand{\tri}{
  \begin{tikzpicture} [scale=0.5]
    \draw (0,0) circle (1) ;
    \draw (0,-1) -- (0,1);
    \draw (-1.5,0) -- (-1,0);
    \draw (0,1) -- (1,1);
    \draw (0,-1) -- (1,-1);
  \end{tikzpicture}
}
\newcommand{\glass}{
  \begin{tikzpicture} [scale=0.5]
    \draw (0,0) circle (1) ;
    \draw (2,0) circle (1) ;
    \draw (-1.5,0) -- (-1,0);
    \draw (3,0) -- (3.5,0);

  \end{tikzpicture}
}

\newcommand{\trianx}{
  \begin{tikzpicture} [scale=0.5]
    \draw (20:1) -- (-20:3);
    \draw[white, line width=4pt] (-20:1) -- (20:3);
    \draw (-0.5,0) -- (0,0);
    \draw (0,0) -- (20:3.5);
    \draw (0,0) -- (-20:3.5);
    \draw (-20:1) -- (20:3);
  \end{tikzpicture}
}

\newcommand{\diag}{
\begin{tikzpicture} [scale=0.5]
\draw (0,-1) -- (0,1);
\draw (0,-1) -- (-0.5,-1.5);
\draw (0,1) -- (-0.5,1.5);
\draw (0,1) -- (2,1);
\draw (0,-1) -- (2,-1);
\draw (2,-1) -- (2,1);
\draw (2,-1) -- (2.5,-1.5);
\draw (2,1) -- (2.5,1.5);
\draw (0,1) -- (2,-1);
\end{tikzpicture}
}

\newcommand{\mug}{
\begin{tikzpicture} [scale=0.5]
\draw plot[smooth] coordinates{(2,1) (2.35,0.5) (2.5,0) (2.35,-0.5) (2,-1)};
\draw (0,-1) -- (0,1);
\draw (0,-1) -- (-0.5,-1.5);
\draw (0,1) -- (-0.5,1.5);
\draw (0,1) -- (2,1);
\draw (0,-1) -- (2,-1);
\draw (2,-1) -- (2,1);
\draw (2,-1) -- (2.5,-1.5);
\draw (2,1) -- (2.5,1.5);
\end{tikzpicture}
}

\newcommand{\ladder}{
\begin{tikzpicture} [scale=0.5]
\draw (0,-1) -- (0,1);
\draw (0,-1) -- (-0.5,-1.5);
\draw (0,1) -- (-0.5,1.5);
\draw (0,1) -- (2,1);
\draw (0,-1) -- (2,-1);
\draw (2,-1) -- (2,1);
\draw (0,1) -- (4,1);
\draw (0,-1) -- (4,-1);
\draw (4,-1) -- (4,1);
\draw (4,-1) -- (4.5,-1.5);
\draw (4,1) -- (4.5,1.5);
\end{tikzpicture}
}

\newcommand{\laddern}{
\begin{tikzpicture} [scale=0.5]
\draw (0,-1) -- (0,1);
\draw (0,-1) -- (-0.5,-1.5);
\draw (0,1) -- (-0.5,1.5);
\draw (0,1) -- (2,1);
\draw (0,-1) -- (2,-1);
\draw[->,line width=1.5pt] (2,-1) -- (2,0);
\draw[line width=1.5pt] (2,0) -- (2,1);
\draw (0,1) -- (4,1);
\draw (0,-1) -- (4,-1);
\draw[->,line width=1.5pt] (4,-1) -- (4,0);
\draw[line width=1.5pt] (4,0) -- (4,1);
\draw (4,-1) -- (4.5,-1.5);
\draw (4,1) -- (4.5,1.5);
\end{tikzpicture}
}

\newcommand{\npladder}{
\begin{tikzpicture} [scale=0.5]
\draw (0,-1) -- (0,1);
\draw (0,-1) -- (-0.5,-1.5);
\draw (0,1) -- (-0.5,1.5);
\draw (0,1) -- (2,1);
\draw (0,-1) -- (2,-1);
\draw (2,-1) -- (4,1);
\draw (0,1) -- (4,1);
\draw (0,-1) -- (4,-1);
\draw (4,-1) -- (3.1,-0.1);
\draw (2,1) -- (2.9,0.1);
\draw (4,-1) -- (4.5,-1.5);
\draw (4,1) -- (4.5,1.5);
\end{tikzpicture}
}

\newcommand{\npladdertwo}{
\begin{tikzpicture} [scale=0.5]
\draw (0,-1) -- (0,1);
\draw (0,-1) -- (-0.5,-1.5);
\draw (0,1) -- (-0.5,1.5);
\draw (0,1) -- (2,1);
\draw (0,-1) -- (2,-1);
\draw (2,-1) -- (4,1);
\draw (0,1) -- (4,1);
\draw (0,-1) -- (4,-1);
\draw (4,-1) -- (3.1,-0.1);
\draw (2,1) -- (2.9,0.1);
\draw (4,-1) -- (4.5,-1.5);
\draw (4,1) -- (4.5,1.5);
\draw[fill=black] (0,0) circle (0.2);
\end{tikzpicture}
}

\newcommand{\low}[1]{\mathbf{#1^-}}

\begin{document}

\thispagestyle{empty}

\vspace{ -3cm} \thispagestyle{empty} \vspace{-1cm}
\begin{flushright} 
\footnotesize
HU-EP-13/68\\
\end{flushright}%

\begingroup\centering
{\Large\bfseries\mathversion{bold}
Non-planarity through unitarity in ABJM
\par}%
\vspace{7mm}

\begingroup
Lorenzo~Bianchi, Marco S. Bianchi\\
\endgroup
\vspace{8mm}
\begingroup\small
\emph{Institut f\"ur Physik,
Humboldt-Universit\"at zu Berlin,
Newtonstra{\ss}e 15, 12489 Berlin, Germany}\\
\endgroup

\vspace{0.3cm}
\begingroup\small
 {\tt $\{$lorenzo., marco.$\}$bianchi@\,physik.hu-berlin.de}
\endgroup
\vspace{1.0cm}

\textbf{Abstract}\vspace{5mm}\par
\begin{minipage}{14.7cm}

We use unitarity techniques to compute the two-loop non-planar corrections to the Sudakov form factor and the four-point amplitude in ABJM theory. 
We start by reconstructing non-planar integrals from two-particle cuts in three dimensions. This causes ambiguities, due to the one-loop four-point amplitude being subleading in dimensional regularization.
We provide a prescription to circumvent them
and show that it leads to the correct results, as checked against the recent Feynman diagram computation.
For the amplitude we point out an alternative basis of integrals, including a non-planar double-box with a numerator inspired by color-kinematics duality. We reproduce the result using a combination thereof with the coefficients fixed by generalized unitarity.
For BLG theory we propose that this gives the form of the amplitude satisfying color-kinematics duality.
Finally, we compute the complete two-loop amplitude of three-dimensional ${\cal N}=8$ SYM, and the corresponding four-point amplitude in ${\cal N}=16$ supergravity as a double copy.

\end{minipage}\par
\endgroup

\section{Introduction}

In this paper we use unitarity techniques to compute the two-loop color subleading corrections to the Sudakov form factor and the four-point amplitude in ABJM theory \cite{ABJM}.

Unitarity aims at determining loop corrections to physical quantities, such as scattering amplitudes, by using on-shell data only, in order to circumvent cumbersome Feynman diagram computations. 
In its original application \cite{Bern:1994zx,BDK} it was based on the Cutkosky rules \cite{Cutkosky:1960sp}, and prescribed how to reconstruct an amplitude from the analysis of its discontinuities in the complex space of the Lorentz invariants it depends on.

Under the assumption that the result is given by a combination of Feynman integrals with rational coefficients, one is able to fix them by comparing the cuts of the amplitude, separating it into lower order on-shell pieces, with the cuts of the integrals. 

The basis of integrals can be in principle reconstructed by inspection of some cuts of the amplitude, by uplifting them to proper Feynman integrals and checking that all other cuts are then satisfied.

Alternatively, one can make an ansatz for the basis of integrals and fix their coefficients via cuts. An efficient strategy to carry out this task consists in performing several cuts on the same diagram, a method referred to as generalized unitarity \cite{Bern:1997sc,BCF2}\footnote{The term was first mentioned in \cite{Eden:98637}}.
This route has been particularly effective for scattering amplitudes of four-dimensional maximally supersymmetric Yang-Mills theory, where in the planar limit such a basis is highly constrained by dual conformal invariance \cite{Drummond:2006rz}. 
The computational efficiency of unitarity triggered spectacular advances in the determination of amplitudes in ${\cal N}=4$ SYM in four dimensions, such as the calculation of the complete four- \cite{Bern:2010tq} and five-loop \cite{Bern:2012uc} four-point amplitudes.

Remarkably, unitarity cuts have been applied also to other quantities such as form factors \cite{Brandhuber:2010ad,Bork:2010wf,Brandhuber:2011tv,Bork:2011cj,Brandhuber:2012vm} and correlation functions \cite{Engelund:2012re} in ${\cal N}=4$ SYM and to supersymmetric theories in different dimensions. It has been used in five dimensions for testing the finiteness of maximally supersymmetric Yang-Mills theory at six loops \cite{Bern:2012di}, in three-dimensional ${\cal N}=6$ superconformal Chern-Simons theories, determining the planar two-loop amplitudes at four \cite{CH} and six points \cite{CaronHuot:2012hr}, and also in two-dimensional models \cite{Engelund:2013fja,Bianchi:2013nra}.

Away from the large $N$ approximation or for observables which do not possess dual conformal invariance such as form factors, a powerful tool for determining the non-planar integral numerators has been offered by the BCJ relations \cite{Bern:2008qj}.
These are kinematic identities among tree level amplitudes relying on the possibility to rewrite them in such a way that their kinematic parts obey the same Jacobi identities as their color factors. This relates tree level color ordered partial amplitudes reducing the number of independent ones.
At loop level unitarity allows to construct amplitudes by fusing tree level ones. In this way the properties of tree level amplitudes propagate to higher loops. The application of the BCJ relations to tree sub-graphs provides identities between the numerators of different loop integrals. Interestingly, they are able to relate the numerators of planar and non-planar topologies, highly constraining the form of the latter \cite{Bern:2008qj}.

It has been conjectured that at both tree and loop level one can express Yang-Mills amplitudes in a form where a color-kinematics duality is satisfied. This happens to be a very general statement, regardless of supersymmetry and dimension \cite{Bern:2008qj}.
Furthermore it has been recently shown that BCJ identities also apply to form factors \cite{Boels:2012ew}.

It was conjectured \cite{Bern:2008qj,Bern:2010ue} and proven at tree level in \cite{Bern:2010yg} that, whenever a gauge theory amplitude is expressed this way, the corresponding gravity amplitude can be obtained as a double copy, namely replacing the color coefficients by another power of the kinematic factor, satisfying the duality. It has also been checked that the structure of IR divergences in gauge and gravity theories are compatible with the double copy procedure \cite{Oxburgh:2012zr}.
For loop amplitudes this leads to an improved ultraviolet behaviour of supergravity theories \cite{Bern:2007hh,Bern:2010ue}. This is particularly interesting for the ${\cal N}=8$ maximal supergravity which has been verified to be perturbatively finite up to four loops \cite{Bern:2008pv,Bern:2009kd}.
Recently it has been shown that half-maximally supersymmetric supergravity in four dimensions, though displaying nicer ultraviolet properties than expected, is divergent at four loops \cite{Bern:2013uka}.

In three dimensions color-kinematics duality and the double copy procedure also apply to Yang-Mills theory and in particular to the maximally supersymmetric ${\cal N}=8$ model.
Another interesting class of field theories, namely the superconformal ${\cal N}=6$ Chern-Simons models, has a very different nature. Indeed, the scattered particles transform in the bi-fundamental representation of the gauge groups and the underlying color structure can be thought of as a three-algebra, rather than a conventional Lie algebra \cite{BL1,BL2,G}.
Nevertheless it has been shown that BCJ relations exist in these models \cite{Bargheer:2012gv}.
In particular, a recent analysis demonstrated that the maximally supersymmetric ${\cal N}=8$ BLG theory \cite{BL1,BL2,G} possesses BCJ identities for any multiplicity, whereas they exist in ABJM only up to six points \cite{Huang:2013kca}.
Focussing on BLG theory, it has been pointed out that reformulating tree level amplitudes in a fashion satisfying color-kinematics duality one can reproduce ${\cal N}=16$ supergravity amplitudes via a double copy \cite{Bargheer:2012gv,Huang:2012wr}.
Since amplitudes in ${\cal N}=8$ SYM are supposed to square to those of ${\cal N}=16$ supergravity as well, this gives two equivalent double copy gravity amplitudes, based on two- and three-algebras, respectively \cite{Huang:2012wr}. 

This intriguing equivalence and the possibility that three-dimensional supergravity theories, which are also power counting nonrenormalizable as in four dimensions, can enjoy an improved ultraviolet behaviour motivate the problem of computing subleading contributions to loop amplitudes in ${\cal N}=6$ Chern-Simons matter theories. 

\bigskip
In this paper we take the first steps addressing this task, using unitarity.
We first tackle the problem of finding non-planar integrals by a constructive strategy, which aims at determining the integrals contributing to the computation, through the analysis of two-particle cuts.
Such a program has been successfully carried out for the color leading two-loop corrections to the form factor in \cite{Brandhuber:2013gda}.
However the application of this approach to subleading contributions reveals subtleties. 
Indeed, using the recently proposed one-loop non-planar amplitudes \cite{Brandhuber:2013gda} in the cuts produces two-loop integrals whose propagators do not correspond to any Feynman diagram. 
We argue that the appearance of such unphysical integrals can be traced back to our loose treatment of the one-loop amplitude in unitarity computations. Namely we perform cuts in three dimensions, while such an amplitude is subleading in the dimensional regularization parameter.
We put forward a strategy to circumvent this obstacle that we briefly outline as follows.

The complete one-loop amplitude in ABJM was constructed by the analysis of its cuts and is expressed in terms of combinations of a dual conformally invariant box function, which is of order $\epsilon$ in dimensional regularization.

However we find that a cut analysis in strictly three dimensions does not give a unique answer in terms of box functions. Rather, the form of the cuts in the $s$, $t$ and $u$ channels can be always reproduced either by one single box function or by the difference of two.
Even though these objects have the same three-dimensional cuts they are not identical.
Indeed they are only equivalent up to the first nontrivial order, when expanded in the dimensional regularization parameter, and up to a constant imaginary term.
Such an identity and its failure to hold beyond leading order in $\epsilon$ are demonstrated by computing the relevant one-loop dual conformally invariant integrals to all orders.

Since such amplitudes are all of order $\epsilon$ this discrepancy does not pose any problems for the one-loop amplitude itself, but can and does generate ambiguities whenever the one-loop amplitude is used as an input in a higher loop computation. This is not surprising and is an artefact of such contributions being subleading in $\epsilon$. 
 In order to fix the correct form of the one-loop amplitude, respecting cuts to higher order in $\epsilon$, we perform cuts in $d$ dimensions, which unambiguously fix the form of the amplitude at one loop.
At two loops, we show that use of three-dimensional cuts is again subtle. Indeed we verify that they can lead to inconsistent Feynman integrals with spurious propagators. This problem could be solved considering $d$ dimensional cuts throughout the whole computation, which would however make it more cumbersome.
 Alternatively, we propose a recipe how to circumvent such subtleties, avoiding the need for complicated $d$-dimensional cuts. Namely, instead of using an explicit form for the one-loop amplitude we just exploit its color structure, its cuts  in three dimensions and the fact that it is expressed in terms of a combination of dual conformally invariant box functions. As recalled above, there are two such combinations.
Depending on which of the two we use to reconstruct the correct two-particle cuts, we always find that one choice gives an unphysical result. Namely, the cut is reproduced by an inconsistent integral with a spurious propagator. Our recipe simply consists of choosing the other combination, which yields well-defined Feynman integrals.
We find that this choice always coincides with a form where the cuts of the one-loop part are reproduced by a single box function.

\bigskip
We first apply this strategy to the computation of the two-loop Sudakov form factor.
We show that the leading and subleading computations are mapped into each other, in such a way that the non-planar contribution turns out to be opposite to the large $N$ one. This was first seen from the Feynman diagram computation of \cite{Bianchi:2013iha}.
We stress that this implies the appearance of a non-planar correction to the cusp anomalous dimension of ABJM already at leading order in the weak coupling expansion.
This is rather different compared to ${\cal N}=4$ SYM, where no nonplanar contribution to the form factor appears up to three loops.

The computation of the two-loop amplitude presents the same subtleties.
If one tries to determine the kind of integrals contributing to it from the two-particle cuts, using the one-loop amplitude in the form given in \cite{Brandhuber:2013gda}, one finds the planar double-box integrals but also unphysical non-planar topologies with spurious propagators.
Again we show that this problem can be circumvented, as outlined above. In this way only proper Feynman integrals are generated.
Interestingly, by exploiting the symmetry properties of the one-loop integrals, no non-planar double-box topology is required for the computation. Rather, the non-planar sector may be expressed in terms of simpler integrals depending on one scale only, which happen to be maximally transcendental \cite{Kotikov:2001sc,Kotikov:2004er}.
The two-particle cuts are not sufficient to fix a combination thereof completely, since they could miss integrals which vanish in the cut configurations.
We then complete our integral combination by imposing the vanishing of three-particle cuts.
The combination we obtain is consistent with quadruple cuts, which separate the two-loop amplitude into three four-point tree level ones. Such a cut does not suffer from the ambiguities due to the one-loop amplitude of order $\epsilon$ and therefore is a nontrivial check that the answer is correct.
Indeed we verify that it reproduces the known result from the Feynman diagram computation \cite{Bianchi:2013iha}. This constitutes the main test that our strategy succeeds in giving the correct result.

\bigskip
Instead of constructing the relevant integrals from the cuts, one can formulate a reasonable ansatz on the basis of integrals appearing in the result and check that their cuts be compatible with those of the amplitude to determine the suitable combination.
Guided by the dual conformally invariant planar integrals and their numerators, we propose a natural non-planar counterpart. This includes a non-planar double-box, whose numerator we guess by a sort of BCJ identity on a particular cut isolating a four-point amplitude.
Specifically, starting from the planar double-box topology of the four-point two-loop color leading amplitude, we cut the diagram in such a way to isolate a four-point tree sub-amplitude. On this we perform a change of two external legs turning the planar topology into a non-planar one. From the antisymmetry property of the BLG four-point amplitude, we conjecture a consistent form of the non-planar numerator.
Next we fix the coefficients of these integrals by demanding that the two- and three-particle cuts are satisfied.

The non-planar double-box presents a complicated numerator, which we deal with by reduction to master integrals through integration by parts identities.
We solve the relevant master integrals by writing down their Mellin-Barnes representation and by repeated use of the Barnes lemmas and their corollaries. 
Finally, assembling the result, we are again able to reproduce the form of the ABJM and BLG four-point amplitudes at two loops.

For the BLG theory we propose that this form is the one respecting color-kinematics duality. Indeed the numerators of the planar and nonplanar integrals are connected by an identity derived from the color properties of a four-point sub-amplitude, which can be isolated by suitable cuts. Properly squaring the numerators should give the corresponding two-loop four-point ${\cal N}=16$ supergravity amplitude.

\bigskip
We next compute the the two-loop non-planar amplitude of three-dimensional ${\cal N}=8$ SYM. This is governed by the non-planar scalar double-box in three dimensions, which appears among the master integrals of the ABJM calculation. We show that, as happens in ${\cal N}=4$ SYM, the subleading single trace partial amplitudes exhibit softer infrared divergences than the leading contribution.

We then consider the form of the complete amplitude satisfying color-kinematics duality and compute the ${\cal N}=16$ supergravity amplitude as its double copy.
In \cite{Huang:2012wr} an equivalence between tree level double copy amplitudes of BLG and ${\cal N}=8$ SYM was pointed out. It would be interesting to check if squaring the BLG amplitude we propose reproduces the four-point two-loop amplitude we derive from ${\cal N}=8$ SYM. This would constitute a nontrivial check at loop level of the equivalence suggested in \cite{Huang:2012wr}.

\bigskip
The plan of the paper is as follows: in section \ref{sec:1} we review the necessary material which is required for our two-loop unitarity based computations, namely the one-loop amplitude and its properties. Here we also propose the alternative form turning out to be crucial for the application of unitarity in the non-planar regime. In section \ref{sec:2}  we spell out the computation of the subleading corrections to the form factor.
In section \ref{sec:3} we perform a computation of the two-loop subleading four-point amplitude by constructing its integrals from two-particle cuts.
In section \ref{sec:4} we derive an alternative ansatz for the integral basis by deforming the planar topologies to non-planar ones and guessing their numerators. Generalized unitarity is used to fix their coefficients.
Then we use the explicit form of the integrals to correctly reproduce the subleading contribution to the amplitude.
Finally in section \ref{sec:5} we use the scalar double-box to derive an expression for the subleading partial amplitude of ${\cal N}=8$ SYM in three dimensions. Using the double copy prescription we compute the four-point two-loop amplitude in ${\cal N}=16$ supergravity.
Appendices follow collecting our conventions and the details of the reduction to master integrals of the non-planar double-box and their computation.

\section{Review of leading and subleading four-point amplitudes in\\ ABJM}\label{sec:1}

We work in ABJM theory with gauge groups $U(N)_k\times U(N)_{-k}$, where $k$ and $-k$ are the two Chern-Simons levels in the action.
We shall take $k$ to be large in order for a perturbative expansion to be possible and rescale it as $K = 4\pi k$ for convenience.
The physical degrees of freedom of the theory are carried by the matter fields, which transform in the (anti)-bifundamental representation of the gauge groups and can be organized into two ${\cal N}=3$ on-shell superfields $\Phi$ and $\bar \Phi$ \cite{BLM}
\begin{align}
  \Phi (\l, \eta) & =
  \phi^4(\l) + \eta^A \psi_A(\l) + \frac{1} {2} \e_{ABC} \eta^A \eta^B
  \phi^C(\l) + \frac{1}{3!} \e_{ABC} \eta^A \eta^B \eta^C \psi_4(\l) \, ,
  \\
  \bar{\Phi} (\l , \eta) & =
  \bar{\psi}^4(\l) + \eta^A \bar{\phi}_A(\l)
  + \frac{1} {2} \e_{ABC} \eta^A \eta^B \bar{\psi}^C(\l) + \frac{1} {3!}
  \e_{ABC} \eta^A \eta^B \eta^C \bar{\phi}_4(\l)
\end{align}
The former is bosonic whereas the latter is fermionic.
Such a superspace is parametrized by Grassmann variables $\eta_A$ with $A=1,2,3$. 
The momenta of the particles are conveniently expressed in terms of commuting spinors $\lambda_i$. Our conventions are collected in Appendix \ref{app:1} and follow those of \cite{Brandhuber:2012un}.
Component amplitudes can then be compactly organized in superamplitudes for the ${\cal N}=3$ superfields of the form ${\cal A}\left(\bar \Phi_1, \Phi_2, \dots, \Phi_{n} \right)$.
Gauge invariance demands $n$ to be even and therefore amplitudes with an odd number of particles identically vanish.

It proves useful to expand superamplitudes in a basis of independent color structures, spanning a color space, whose coefficients are the color ordered partial amplitudes.
Such a color decomposition is possible for bifundamental superfields and reads \cite{BLM}
\begin{equation}
\tilde{\cal A}_n \left( \bar \Phi^{\bar a_1}_{1 \; a_1},\,  \Phi^{b_2}_{2 \; \bar b_2}, \, \bar \Phi^{\bar a_3}_{3 \; a_3}, \cdots 
\Phi^{b_n}_{n \; \bar b_n} \right) = \left(\frac{4\pi}{K}\right)^{\frac{n}{2}-1}
  \sum_{\s}
\mathcal{A}_n(\s(1),   \cdots , \s(n) ) \; \d^{\bar a_{\s(1)}}_{\bar b_{\s(2)}} \, \d^{b_{\s(2)}}_{a_{\s(3)}}
\cdots \d^{b_{\s(n)}}_{a_{\s(1)}}
\end{equation}
where the sum is over exchanges of even and odd sites among themselves, up to cyclic permutations by two sites.
We refer to color ordered amplitudes with ${\cal A}$ and denote the complete, color dressed amplitudes by $\tilde{\cal A}$.
In particular at four points there exist four possible index contractions among four superfields $ (\bar \Phi_1)^{\bar i_1}_{\ i_1} (\Phi_2)^{i_2}_{\ \bar i_2}  (\bar \Phi_3)^{\bar i_3}_{\ i_3} (\Phi_4)^{i_4}_{\ \bar i_4}$ transforming in the (anti)-bifundamental representations of the $U(N)\times U(N)$ gauge group.

We follow the notation of \cite{Brandhuber:2013gda}, and adopt a compact representation for them with square brackets
\begin{align}
 [1,2,3,4] &= \d^{i_2}_{i_1} \, \d^{\bar i_3}_{\bar i_2}  \, \d^{i_4}_{i_3} \, \d^{\bar i_1}_{\bar i_4} \,  ,& 
[1,4,3,2] &=  \d^{i_4}_{i_1} \, \d^{\bar i_3}_{\bar i_4}  \, \d^{i_2}_{i_3} \, \d^{\bar i_1}_{\bar i_2} \, ,\non \\
[1,2][3,4] &=  \d^{\bar i_1}_{\bar i_2} \, \d^{i_2}_{i_1}  \, \d^{\bar i_3}_{\bar i_4} \, \d^{i_4}_{i_3} \, ,& 
[1,4][3,2] &=   \d^{\bar i_1}_{\bar i_4} \, \d^{i_4}_{i_1}  \, \d^{\bar i_3}_{\bar i_2} \, \d^{i_2}_{i_3} \, .\label{traces}
\end{align}
Color ordered $n$-point amplitudes at $l$ loops in ABJM obey the following symmetry properties
\begin{equation}\label{eq:symm1}
{\cal A}^{(l)} (\bar{1}, 2, \bar{3} \dots, n) = (-1)^{\frac{n}{2} - 1}\, {\cal A}^{(l)} (\bar{3}, 4,  \dots, \bar{1}, 2)
\end{equation}
and
\begin{equation}\label{eq:symm2}
{\cal A}^{(l)} (\bar{1}, 2, \bar{3} \dots, n) =
(-1)^{\frac{n(n-2)}{8} + l} \, {\cal A}^{(l)} (\bar{1}, n, \overline{n-1}, n-2, \dots, \bar{3}, 2)
\end{equation}
For the four-point case this entails that there exists a unique tree level partial amplitude which reads \cite{BLM}
\begin{equation}
{\cal A}^{(0)}_4 (\bar{1}, 2, \bar{3}, 4) = i\, \frac{\delta^{(6)}(Q)\delta^{(3)}(P)}{\braket{1 2}\braket{2 3}}
\end{equation}
where $P_{\alpha\beta}=\sum_{i=1}^4\, \lambda_{i\, \alpha} \lambda_{i\, \beta}$ and $Q^A_\alpha = \sum_{i=1}^4\, \lambda_{i\, \alpha} \eta^{A}_i$.
The bosonic and fermionic $\delta$-functions enforce momentum conservation and supersymmetry, respectively. Component amplitudes can be read off the superamplitude by expanding the fermionic $\delta^{(6)}(Q)$ in $\eta$.
The four-point superamplitude will be used extensively in the unitarity cuts. 

Six-point tree level amplitudes were also computed in \cite{BLM} and they were found to be Yangian \cite{BLM,DHP} and dual superconformal invariant \cite{HL,Drummond:2008vq}.
Via a three-dimensional version of the BCFW relations \cite{BCF,BCFW} determined in \cite{GHKLL} it is then possible to derive all tree level amplitudes, which will also be automatically Yangian invariant.

Notably, tree level ABJM amplitudes and the leading singularities of loop ones can be formulated in terms of an orthogonal Grassmannian integral formalism \cite{Lee}, similarly to the ${\cal N}=4$ SYM case \cite{ArkaniHamed:2009dn,ArkaniHamed}.
A proposal to extend the on-shell diagram description of amplitudes in ${\cal N}=4$ SYM \cite{ArkaniHamed:2012nw} to ABJM has been recently put forward \cite{Huang:2013owa}.

\subsection{The one-loop function}

The one-loop four-point amplitudes in ABJM were first studied in \cite{ABM} by a Feynman diagram analysis, which uncovered that they are subleading in $\epsilon$ when using dimensional regularization (the same computation in ${\cal N}=2$ superspace language was performed in \cite{BLMPS2}).
The planar partial amplitudes were given a manifestly dual conformally invariant integral form in \cite{CH}. They are expressed in terms of the one-loop box function
\begin{equation}\label{eq:boxfunction}
I(1,2,3,4) \equiv \int \frac{d^dl}{(2\pi)^d\, i}\, \frac{s\, \Tr(l\, p_1\, p_4) + l^2\, \Tr(p_1\, p_2\, p_4)}{l^2\, (l-p_1)^2\, (l-p_{12})^2\, (l+p_4)^2}
\end{equation}
where $p_{12}=p_1+p_2$.
It consists of a vector box and a scalar triangle, whose combination preserves dual conformal invariance.
Then the one-loop amplitude reads
\begin{equation}
\frac{\tilde{\cal A}^{(1)} (\bar 1,2,\bar 3,4)}{{\cal A}^{(0)} (\bar 1,2,\bar 3,4)}\, \bigg|_{planar}\ =\, i\, N\, \left( [1,2,3,4] + [1,4,3,2] \right)\, I(1,2,3,4)
\end{equation}
The analysis of the $s$- and $t$-channel cuts reveals that this integral behaves correctly under the requirements of unitarity.
Consistently its explicit evaluation reveals that it is subleading in $\epsilon$ when dimensionally regulated in $d=3-2\epsilon$ dimensions. 

In the planar limit six-point one-loop amplitudes have been studied in \cite{Bargheer:2012cp,Bianchi:2012cq,Brandhuber:2012un} and a recursion relation for one-loop supercoefficients was pointed out in \cite{Brandhuber:2012wy}, which allows any one-loop amplitudes in principle to be constructed.

Recently unitarity has been used to express also color subleading contributions to the four-point one-loop amplitude in ABJM \cite{Brandhuber:2013gda}, revealing that these are also governed by combinations of the one-loop dual conformally invariant box function \eqref{eq:boxfunction}.

In the same paper \cite{Brandhuber:2013gda} some useful properties of this function and its cuts were pointed out, which are crucial for determining the amplitude via unitarity and for verifying its consistency with the symmetry property \eqref{eq:symm2}.
Among those we recall its antisymmetry under exchange of the momenta in the odd or even positions and under cyclic permutations
\begin{equation}\label{eq:antisymmetry}
I(a,b,c,d) = - I(c,b,a,d) = -I(a,d,c,b) \qquad I(a,b,c,d) = - I(b,c,d,a)
\end{equation}
Furthermore an identity was pointed out between the symmetrized cuts of the integrals 
\begin{equation}\label{eq:symmcut}
{\cal S}_{12}\ I(1,2,3,4) \big|_{s-cut}\ =\ {\cal S}_{12}\ I(1,2,4,3) \big|_{s-cut}
\end{equation}
These properties, valid strictly in three dimensions (see Appendix \ref{app:dcut}), can be used to prove that for instance the combination
$I(4,2,3,1)-I(1,2,3,4)$ has the correct cuts to reproduce the one-loop four-point partial amplitude associated to the color structure $[1,2][3,4]$.
Henceforth the complete one-loop amplitude reads \cite{Brandhuber:2013gda}
\begin{align}\label{eq:amp1L}
\frac{\tilde{\cal A}^{(1)} (\bar 1,2,\bar 3,4)}{{\cal A}^{(0)} (\bar 1,2,\bar 3,4)} =& \, i\,  (N\, [1,2,3,4] +N\, [1,4,3,2] -2 [1,2][3,4] - 2 [1,4][2,3])\, I(1,2,3,4)
\nonumber\\&
 + 2\, i\, [1,2][3,4]\, I(4,2,3,1) - 2\, i\, [1,4][2,3]\, I(1,3,4,2)
\end{align}
We observe that using the relations (\ref{eq:antisymmetry}) and (\ref{eq:symmcut}) the integral $-I(1,2,4,3)$
also possesses the correct cuts to serve as the one-loop integral for the partial amplitude corresponding to $[1,2][3,4]$. Indeed
\begin{align}\label{eq:cutanalysis}
& -{\cal S}_{12} I(1,2,4,3) \big|_{s-cut} = {\cal S}_{12} \left[ I(4,2,3,1)-I(1,2,3,4) \right] \big|_{s-cut} = -{\cal S}_{12} I(1,2,3,4) \big|_{s-cut} \nonumber\\&
-{\cal S}_{23} I(1,2,4,3) \big|_{t-cut} = 0 = {\cal S}_{23} \left[ I(4,2,3,1)-I(1,2,3,4) \right] \big|_{t-cut} \nonumber\\&
-{\cal S}_{24} I(1,2,4,3) \big|_{u-cut} = {\cal S}_{24} I(4,2,1,3) \big|_{u-cut} = {\cal S}_{24} I(4,2,3,1)\big|_{u-cut} = \nonumber\\&
~~~~~~~~~~~~~~~~~~~~~~~~~~~~~ = {\cal S}_{24} \left[ I(4,2,3,1)-I(1,2,3,4) \right] \big|_{u-cut}
\end{align}
The cuts for the partial amplitude associated to $[1,4][2,3]$ can be analogously reconstructed using $-I(4,2,3,1)$.
Choosing these integrals, the amplitude takes the form
\begin{align}\label{eq:myamp}
\frac{\tilde{\cal A}^{(1)} (\bar 1,2,\bar 3,4)}{{\cal A}^{(0)} (\bar 1,2,\bar 3,4)} =&\,   i\, N\, ( [1,2,3,4] + [1,4,3,2])\,  I(1,2,3,4)
\nonumber\\&
 - 2\,i\,  [1,2][3,4]\, I(1,2,4,3) - 2\,i\,  [1,4][2,3]\, I(4,2,3,1)
\end{align}

In particular it is easy to check from the explicit evaluation of the box functions $I$ that both one-loop amplitudes are consistently of order $\epsilon$ in dimensional regularization.
This is perfectly consistent with the order in the $\epsilon$ expansion at which these results have been derived. Namely we have been working with strictly three-dimensional cuts, i.e. up to order 0 in $\epsilon$. With this precision expressions \eqref{eq:amp1L} and \eqref{eq:myamp} are indeed equivalent.
However we shall show that they are not identical at subleading orders in $\epsilon$.
This difference, although irrelevant at the level of the one-loop amplitude in three dimensions, becomes important in a unitarity based computation which fuses the one-loop amplitude with some other object. This procedure can generate divergences, which translate in $\epsilon$ poles, making subleading terms of the cut amplitudes important. Therefore it is necessary to decide which form of the amplitude correctly reproduces the cuts at any order in $\epsilon$. 

In order to better understand the nature of the difference between \eqref{eq:amp1L} and \eqref{eq:myamp} we solve the one-loop function integral to all orders in the dimensional regularization parameter $\epsilon$. 
By multiplying and dividing \eqref{eq:boxfunction} by $\Tr(p_1 p_2 p_4) = \pm \sqrt{s t u}$ we can write down its decomposition in terms of scalar integrals which reads (taking e.g. the plus sign)
\begin{equation}\label{decomposition}
I(1,2,3,4) = \frac{s\,t}{2\,i\,\sqrt{s\,t\,u}} \left[ -2\,s\, \TRIANGLE(s) -2\,t\, \TRIANGLE(t) + s\,t\, \BOX(s,t) \right]
\end{equation}
The integral
\begin{equation}\label{eq:triangle}
\TRIANGLE(s) = -i\, \frac{\Gamma\left( 3/2+\epsilon \right) \Gamma^2\left( -1/2-\epsilon \right)}{(4\pi)^{3/2-\epsilon} \Gamma\left(-2\epsilon \right)}\, (-s)^{-3/2-\epsilon}
\end{equation}
stands for the scalar triangle with two on-shell legs in the given channel $s$, and 
\begin{align}\label{eq:box}
\BOX(s,t) &= -i\, \frac{\Gamma\left( 1/2+\epsilon \right) \Gamma^2\left( -1/2-\epsilon \right)}{(4\pi)^{3/2-\epsilon} \Gamma\left(-1-2\epsilon \right)\, s\, t}
\left[ (-t)^{-1/2-\epsilon}\, _2F_1 \left(\begin{array}{c} 1, -1/2-\epsilon;\\ 1/2-\epsilon\end{array} \bigg| 1+\frac{t}{s} \right) + 
\right.\nonumber\\&\left. ~~~~ +
(-s)^{-1/2-\epsilon}\, _2F_1 \left(\begin{array}{c} 1, -1/2-\epsilon;\\ 1/2-\epsilon\end{array} \bigg| 1+\frac{s}{t} \right) 
\right]
\end{align}
is the scalar massless box, whose all order in $\epsilon$ form can be derived from the four-dimensional result \cite{Smirnov:2006ry,Smirnov:2004ym} by a shift in the dimensional regularization parameter, $\epsilon \rightarrow \epsilon+1/2$.
We now study the one-loop function $I(1,2,3,4)$ in the Euclidean regime where $s<0$ and $t<0$. Then the formula \eqref{eq:box} holds, provided an analytic continuation is performed in such a way that the integral $I$ is manifestly real.
The cut analysis performed in \eqref{eq:cutanalysis} suggests that a relation may hold among e.g. $I(1,2,3,4)$, $I(1,4,2,3)$ and $I(1,2,4,3)$.
Inspecting the form of the functions appearing in their exact evaluation, one can easily map the arguments of the Gauss functions into one another by textbook transformations.
However establishing a relation among the hypergeometric functions requires setting $\epsilon=0$, namely at order $\epsilon$ for the box integrals.
In this case their result simplifies considerably, giving e.g.
\begin{equation}\label{eq:oneloopeps}
\BOX(s,t) = \frac{\epsilon}{(s t)^{3/2}} \left[\sqrt{u}\, \log \left(\frac{\sqrt{-s}+\sqrt{-t}+\sqrt{u}}{\sqrt{-s}+\sqrt{-t}-\sqrt{u}}\right)-\sqrt{-s}-\sqrt{-t}\right] + {\cal O}(\epsilon^2)
\end{equation}
We analyze such expressions in the regime where $s$ and $t$ are negative. Since we have used the momentum conservation condition, we then assume $u>0$. 
Under these assumptions and using \eqref{eq:oneloopeps}, we can prove the following identity holds at order $\epsilon$
\begin{equation}
I(1,2,3,4) = I(1,2,4,3) - I(1,4,2,3) + i\, \frac{\pi}{4}\,\epsilon + {\cal O}(\epsilon^2) \label{ident}
\end{equation}
We note in particular the appearance of a constant imaginary part (due to $u$ being positive), which is invisible to the cut analysis performed in \eqref{eq:cutanalysis}.
We remark that the choice we made on the sign of $u$ amounts to inspecting the cut of the various integrals in the $u$-channel, which should be signalled by the appearance of an imaginary part.
In particular, as expected, the integral $I(1,2,3,4)$ has a vanishing cut in the $u$-channel. Indeed it does not develop an imaginary part when we change the sign of $u$, since both the $\log$ and the square root in front of it are purely imaginary and, multiplied, evaluate to a real contribution. 
On the other hand the difference $I(1,2,4,3) - I(1,4,2,3)$ possesses an imaginary part which indicates a discontinuity in the $u$-plane. 
 However, since the cut analysis was performed in strictly three dimensions, this term was overlooked.
In order to achieve a full control on the discontinuities of the amplitude at order $\epsilon$ it is necessary to consider $d$-dimensionsl cuts.
This is performed explicitly in appendix \ref{app:dcut}, where we show how this imaginary part emerges in a $d$-dimensional cut scenario. In this framework we provide evidence that the one-loop amplitude to all orders in $\epsilon$ is that of \eqref{eq:myamp}.

One can feed a higher loop computation with this amplitude via unitarity, but in order to be fully consistent and avoid these kind of ambiguities, one should keep considering $d$-dimensional cuts. 
Indeed, as we shall explicitly show in the next sections, using this amplitude
in two-loop computations with strictly three-dimensional cuts can produce non-physical propagators in the Feynman integrals. This signals that the three-dimensional cut was not able to fully capture the discontinuities of the amplitude and demands for an analysis at higher order in the dimensional regularization parameter. 
Unfortunately the extension of the $d$-dimensional cut technique to higher loops is non-trivial and goes beyond the aim of this work. Rather, here we want to apply a different strategy which, avoiding $d$-dimensional cuts, turns out to be efficient and to correctly reproduce the known two-loop results for amplitudes and form factors.

There are in principle several approaches one can undertake to compute the two-loop integrand.
A first way consists of disregarding two-particle cuts involving the one-loop amplitude and performing multiple cuts until everything is reduced to a product of tree level amplitudes. On the one hand this doesn't involve the one-loop amplitude, and actually such a strategy proves convenient, whenever a basis of integrals is known (such as for the planar two-loop amplitude).
On the other hand it is more difficult to reconstruct two-loop integrals uplifting their quadruple cuts since too many internal momenta have been put on-shell. Additionally, three-particle cuts identically vanish, since they separate the two-loop amplitude into two five-point tree level ones which do not exist in the ABJM theory. Therefore they can only provide a nontrivial check of the relative coefficients of a set of integrals, but do not give any hints about their numerators.

In what follows we shall take an alternative point of view. Namely, we shall work with cuts in strictly three dimensions and consider a generic one-loop amplitude
\begin{align}\label{eq:testamp}
\frac{\tilde{\cal A}^{(1)} (\bar 1,2,\bar 3,4)}{{\cal A}^{(0)} (\bar 1,2,\bar 3,4)} &=\, i\, N\, ( [1,2,3,4] + [1,4,3,2])\,  I(1,2,3,4)
\nonumber\\&
 - 2\,i\, [1,2][3,4]\, A(1,2,4,3) - 2\,i\, [1,4][2,3]\, A(4,2,3,1)
\end{align}
where the object $A(1,2,4,3)$ is defined to be a combination of dual conformally invariant one-loop functions $I$, having the required three-dimensional cuts to reproduce the corresponding subleading partial amplitude. 
In other words $A(1,2,4,3)$ can be either $I(1,2,4,3)$ or $I(1,2,3,4)-I(2,4,1,3)$, and therefore inherits the same symmetry properties (for example antisymmetry under $p_1 \leftrightarrow p_4$ in this case).
As we will see, depending on the cut, one of the two choices is able to produce a physical integral, whereas the other one is not. Our prescription consists of choosing the physical one, plugging it into the cut and reconstructing the corresponding integral from this.

In particular, after performing the first two-particle cut, we know in which channel the one-loop object we have isolated does not possess a cut.
As stated above, this condition can be verified by just one box function or a combination of two.
These are conveniently selected by inspecting which are antisymmetric in the momenta of the given channel.
As an example, both $I(1,2,4,3)$ and $I(1,2,3,4)-I(2,4,1,3)$ have vanishing three-dimensional $t$-channel cut, since they are antisymmetric in $p_1 \leftrightarrow p_4$.
Then, as a rule of thumb, we observe that the physical choice is always to use one single one-loop integral instead of the difference of two\footnote{This choice is supported also by the $d$-dimensional analysis of appendix \ref{app:dcut}.}.
Since we have seen that they differ at order $\epsilon$ by an imaginary part, we expect a spurious imaginary contribution in the two-loop integrals also. We will observe that this is indeed the case, by explicitly computing the unphysical integral arising in the calculation of the subleading form factor.

Having gained an insight into the integrals appearing in the two-loop computation by the method outlined above, we then verify that they correctly satisfy the requirements from generalized quadruple and triple cuts, which are safely free from ambiguities. 

Since the objects $A$ are only defined through their cuts, our procedure might look like performing a quadruple cut. However our prescription contains more information, because we also assume that the additional cut uplifts to a one-loop box function. This puts crucial constraints on the numerator of the two-loop integral, which a quadruple cut would overlook. 

\subsection{The two-loop amplitude}

The two-loop four-point amplitude was determined in the planar limit from unitarity and by   Feynman supergraphs in \cite{CH} and \cite{BLMPS1} respectively.
The color structure of the planar two-loop amplitude is the same as the tree level one, hence we can define the ratio between the complete amplitudes
\begin{equation}
\mathcal{M}_4 \equiv \frac{\tilde{\cal A}^{(2)} (\bar 1,2,\bar 3,4)}{\tilde{\cal A}^{(0)} (\bar 1,2,\bar 3,4)}
\end{equation}
The unitarity computation was performed by setting a basis of dual conformally invariant integrals and then fixing their coefficients imposing double two-particle and three-particle cuts.
The outcome of this analysis is that the two-loop amplitude is correctly reproduced by the combination\footnote{The relative sign between the integrals is different from previous results that have appeared in the literature, due to different conventions.} 
\begin{equation}\label{eq:planarratio}
\mathcal{M}_4 \Big|_{planar} = \textup{\bf{DB}}_P(s,t)+\textup{\bf{DB}}_P(t,s)+\textup{\bf{DT}}_P(s)+\textup{\bf{DT}}_P(t)
 \equiv \textup{\bf{I}}_P(s,t)+\textup{\bf{I}}_P(t,s)
\end{equation}
where we have suppressed the coupling constant.
The integral
\begin{equation}\label{eq:dp}
\textup{\bf{DB}}_P(s,t)=\int\frac{d^dl}{(2\pi)^d}\frac{d^dk}{(2\pi)^d}\frac{\left[s \Tr(l\,p_1\,p_4)+l^2\Tr(p_1\,p_2\,p_4)\right]\left[s\Tr(k\,p_1\,p_4)+k^2 \Tr(p_1\,p_2\,p_4)\right]}{t\,l^2(l+p_3+p_4)^2(l+p_4)^2(k-l)^2k^2 (k-p_1-p_2)^2(k-p_1)^2}
\end{equation}
is a planar double-box whose numerator was shown to be  dual conformally invariant, with a five-dimensional analysis \cite{CH}. 
Expanding the traces this can be decomposed into scalar integrals, which are also dual conformally invariant \cite{CH,BLMPS2}.
Among these there appears the double-triangle 
\begin{equation}
\textup{\bf{DT}}_P(s) =\int\frac{d^dl}{(2\pi)^d}\frac{d^dk}{(2\pi)^d}\frac{s^2}{l^2(l+p_3+p_4)^2(k-l)^2k^2(k-p_1-p_2)^2}
\end{equation}
contributing to \eqref{eq:planarratio}.
We observe that the sum $\textup{\bf DB}_P(s,t) + \textup{\bf DT}_P(s)$ not only satisfies the cuts, but also entails a pleasing cancellation of unphysical infrared divergences which plague the two integrals separately even with off-shell external momenta. This leaves only physical infrared divergences in the amplitude, descending from the massless nature of the scattered particles \cite{BLMPS2}.
Moreover we note that the object $\textup{\bf{I}}_P(s,t)$ is maximally transcendental, since $\textup{\bf DT}_P(s)$ exactly cancels the lower transcendentality parts of $\textup{\bf DB}_P(s,t)$ as can be easily ascertained from their explicit expressions \eqref{db} and \eqref{dt}
\begin{equation}
\textup{\bf{I}}_P(s,t) = -\frac{1}{16\pi^2}\left(\frac{s e^{\g_E}}{4\pi\mu^2}\right)^{-2\e}\left(\frac{1}{(2\e)^2}+\frac{\log2 + \tfrac12 \log\frac{s}{t}}{2\e}-2\z_2 - \log^2 2+\mathcal{O}(\e)\right)\label{maxtrres}
\end{equation}
Using such results the amplitude takes the explicit form
\begin{equation}
\label{eq:planarresult}
{\cal M}_4 \Big|_{planar} = \left(\frac{N}{K}\right)^2 \left(-\frac{(-s/\m'^2)^{-2 \epsilon }+(-t/\m'^2)^{-2 \epsilon }}{(2 \epsilon)^2} + \frac{1}{2} \log ^2 \frac{s}{t} + \frac{2 \pi ^2}{3} + 3 \log^2 2\right)
\end{equation}
where the dimensional regularization mass scale has been redefined as $\mu'^{2}= 8 \pi e^{-\gamma_E}\mu^2$.
This expression is surprisingly similar to the one-loop ${\cal N}=4$ SYM result \cite{CH,BLMPS1,Bianchi:2011aa} and matches the four-cusped Wilson loop computation \cite{HPW,Bianchi:2013pva,BLMPRS}, hinting at a possible amplitude/WL duality \cite{Drummond:2007aua,Drummond:2007cf,Drummond:2007au,Brandhuber:2007yx} in three dimensions.
In the planar limit the two-loop six-point amplitude has also been computed \cite{CaronHuot:2012hr}.

The color subleading corrections to the four-point amplitude were first derived in \cite{Bianchi:2013iha} by a Feynman diagram computation.
Using ${\cal N}=2$ superspace formalism and a clever choice of the external particles makes the evaluation of such contributions feasible. In fact it boils down to summing over permutations of the external legs in the set of graphs already contributing to the leading amplitude, with the addition of just one genuinely non-planar diagram. The corresponding non-planar integral turns out to be easily computable, since in particular it depends on a single momentum invariant. 

The color analysis reveals that only single trace partial amplitudes appear at two loops,  in the same antisymmetric fashion $[1,2,3,4]-[1,4,3,2]$, as for the tree level amplitude.
Therefore one can consider a ratio between the complete two-loop and tree level amplitudes giving the finite $N$ result
\begin{align}
\label{eq:result}
{\cal M}_4 &= \left(\frac{N}{K}\right)^2 \left(-\frac{(-s/\m'^2)^{-2 \epsilon }+(-t/\m'^2)^{-2 \epsilon }}{(2 \epsilon)^2} + \frac{1}{2} \log ^2 \frac{s}{t} + \frac{2 \pi ^2}{3} + 3 \log^2 2\right) + \\ \nonumber
& \! + \frac{1}{K^2} \left(2\, \frac{(-s/\m'^2)^{-2 \epsilon } + (-t/\m'^2)^{-2 \epsilon } - (-u/\m'^2)^{-2 \epsilon }}{(2\epsilon) ^2} + 2 \log \frac{s}{u} \log \frac{t}{u} + \frac{\pi ^2}{3} - 3 \log ^2 2\right)
\end{align}
which is still maximally transcendental, given that the additional non-planar integrals are.
In the rest of the paper we shall reproduce this result from a cut analysis.
To do this we shall start with a simpler case, namely the subleading corrections to the Sudakov form factor.

\section{The color subleading Sudakov form factor from unitarity}\label{sec:2}

Following the computation of \cite{Brandhuber:2013gda}, we determine the subleading correction to the Sudakov form factor in ABJM for a scalar bilinear 1/2-BPS operator. 
As in \cite{Brandhuber:2013gda}, we start from a two-particle cut in three dimensions, dividing the two-loop form factor into
a tree level one and a four-point one-loop amplitude. 

Since we choose a particular projection of the 1/2-BPS operator, we work with particular component amplitudes, which can be extracted from the corresponding superamplitude selecting the correct coefficient of its $\eta$ expansion.
This procedure was successfully applied for the color leading part, as checked against explicit Feynman diagram computations \cite{Young:2013hda,Bianchi:2013iha}.

\begin{figure}[htbp]
\begin{center}
\includegraphics[width=8. cm]{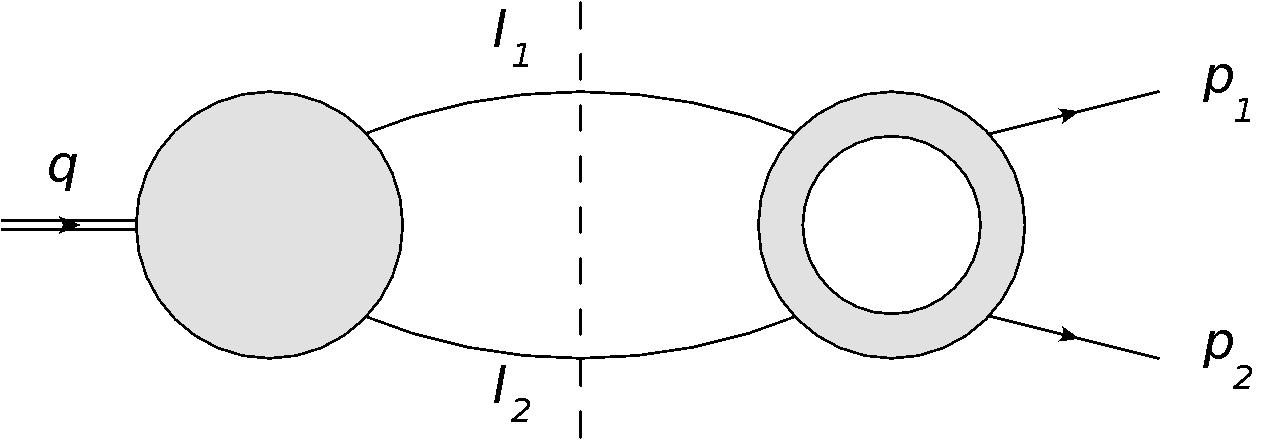}
\caption{Two-particle cut of the Sudakov form factor.}
\label{2ptcutFF}
\end{center}
\end{figure}

As we mentioned in section \ref{sec:1} there are two different ways to obtain the correct cuts for the one-loop amplitude (restricting to cuts in three dimensions): \eqref{eq:amp1L} and \eqref{eq:myamp}. Therefore the two-particle cut shown in Figure \ref{2ptcutFF} contains an ambiguity given by the choice of the expression for the subleading one-loop partial amplitude on the r.h.s. of the cut. Analyzing in detail the example of the form factor, we shall provide a prescription to solve this ambiguity and show that a different method gives an unphysical contribution coming from the imaginary part contained on the r.h.s. of equation \eqref{ident}.

\begin{figure}[htbp]
\begin{center}
\includegraphics[width=15 cm]{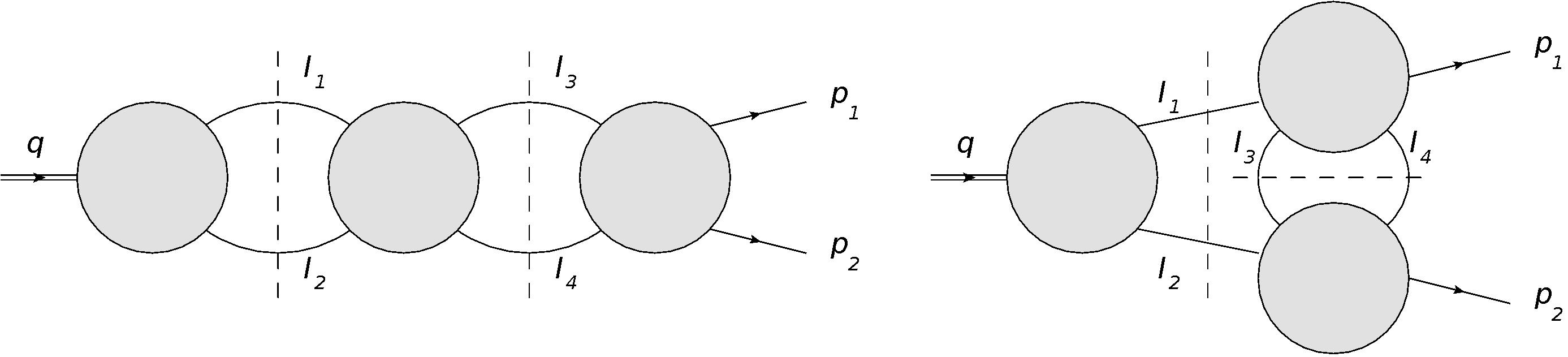}
\caption{Double cuts of the Sudakov form factor.}
\label{doublecutFF}
\end{center}
\end{figure}

To get a deeper insight into the kind of diagrams which can appear in the expression of the two-loop form factor, let us first consider the color factors associated to the quadruple cuts shown in Figure \ref{doublecutFF}. For the first case, combining the color structures of the form factor and the tree level amplitudes one gets
\begin{equation}\label{colorP}
{\cal C}_{P}= [l_2,l_1]([l_1,l_2,l_3,l_4]-[l_1,l_4,l_3,l_2])([l_3,l_4,2,1]-[l_3,1,2,l_4])=0
\end{equation}
This clearly shows that this quadruple cut vanishes and we can't expect any two-loop integral admitting this kind of cut. This strikingly contrasts with SYM theory in four dimensions, due to the different color flux of the diagrams.

In the second double cut the color factor reads
\begin{equation}\label{colorNP}
{\cal C}_{NP}=[l_2,l_1]([l_1,l_3,2,l_4]-[l_1,l_4,2,l_3])([l_3,1,l_4,l_2]-[l_3,l_2,l_4,1])=-2(N^2-1)[1,2]
\end{equation}

\begin{figure}[htbp]
\begin{center}
\includegraphics[width=5 cm]{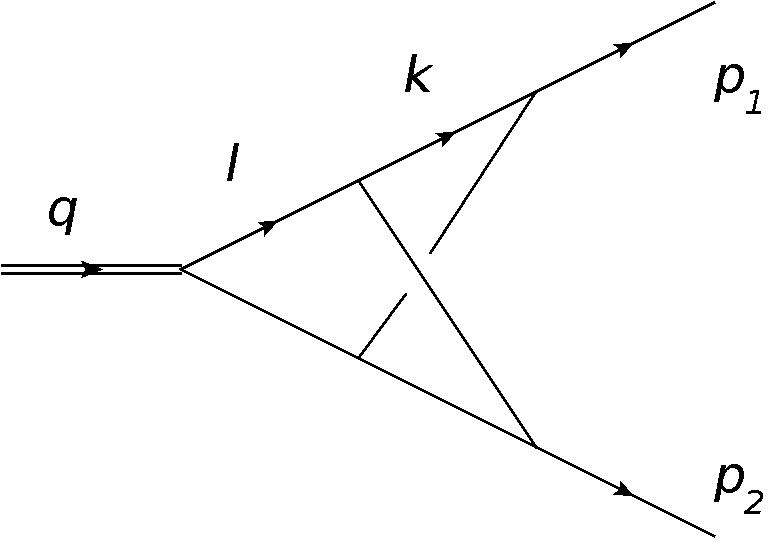}
\caption{Diagram $\textup{\bf{XT}}(q^2)$.}
\label{fig:XT}
\end{center}
\end{figure}

As pointed out already for ${\cal N}=4$ SYM in \cite{Brandhuber:2010ad} and then for ABJM in \cite{Brandhuber:2013gda}, non-planar topologies contribute to the color leading part of the form factor. 
In particular for ABJM the color leading part was shown to be given by a single non-planar maximally transcendental integral
\begin{equation}
F^{(2)}(\bar{p}_1,p_2)=\frac{N^2}{k^2}\, \textup{\bf{XT}}(q^2)
\end{equation}
The integral $\textup{\bf{XT}}$ is shown in Figure \ref{fig:XT} and its explicit expression is
\begin{equation}\label{XT}
\textup{\bf{XT}}(q^2)=\int\frac{d^dl}{(2\pi)^d}\frac{d^dk}{(2\pi)^d}\, \frac{q^2\big(\Tr(p_1\,p_2\,l\,k)-q^2 k^2\big)}{l^2(l-q)^2k^2(l-k)^2(l-k-p_2)^2(k-p_1)^2}
\end{equation}
where $q^2=(p_1+p_2)^2$.

Using the cut in Figure \ref{2ptcutFF} and the information coming from the color factors \eqref{colorP} and \eqref{colorNP} we compute the color subleading part of the two-loop form factor. First of all let us note that, since the first cut in Figure \ref{doublecutFF} is identically vanishing, on the r.h.s. of the cut in Figure \ref{2ptcutFF} we can't expect any one-loop integral having a cut in the channel $q^2$. As outlined in section \ref{sec:1}, we use the additional information that the one-loop amplitude is expressed just in terms of the box function \eqref{eq:boxfunction} and conclude that only two possible combinations can appear on the r.h.s. of the two-particle cut. Indeed only $I(1,-l_1,2,-l_2)$ and $I(1,2,-l_1,-l_2)-I(1,2,-l_2,-l_1)$ have no cut in the $q^2$ channel. With this restriction in mind we evaluate explicitly the two-particle cut with the two expressions of the one-loop amplitude given in section \ref{sec:1}. The integrand is given by
\begin{equation}
F^{(2)}(\bar{p}_1,p_2)\Big|_{\textup{$q^2$-cut}}=F^{(0)}(\bar{l_2},l_1)\,[l_2,l_1]\,\tilde{\cal{A}}_4^{(1)}(\bar{\phi}_A(p_1),\phi^4(p_2),\bar{\phi}_4(-l_1),\phi^A(-l_2))
\end{equation}
Using the form \eqref{eq:amp1L} for the one-loop amplitude we obtain
\begin{equation}\label{FFcut1}
F^{(2)}(\bar{p}_1,p_2)\Big|_{\textup{$q^2$-cut}} = i\, \frac{\braket{12}\braket{1l_1}}{\braket{2l_1}}\, \Big[N^2 \,I(1,-l_1,2,-l_2)-I(1,2,-l_1,-l_2)+I(1,2,-l_2,-l_1)\Big]
\end{equation}
In this expression we note the appearance of the integrals expected from the quadruple cut restrictions. In particular the color leading part is given by a single one-loop integral, while the subleading part is given by a difference which respects, as pointed out in section \ref{sec:1}, the symmetries and the cut requirements.

Using expression \eqref{eq:myamp} for the one-loop amplitude the result is instead
\begin{equation}\label{FFcut2}
F^{(2)}(\bar{p}_1,p_2)\Big|_{\textup{$q^2$-cut}} = i\, \frac{\braket{12}\braket{1l_1}}{\braket{2l_1}}\Big[N^2 \big(I(1,2,-l_1,-l_2)-I(1,2,-l_2,-l_1)\big) +I(1,-l_1,2,-l_2)\Big]
\end{equation}
It is easy to note that the leading and subleading parts of the form factor are interchanged between equation \eqref{FFcut1} and \eqref{FFcut2}. Stated differently, the ambiguity in the calculation of the one-loop amplitude is translated into the possibility of exchanging $I(1,2,-l_1,-l_2)-I(1,2,-l_2,-l_1)$ and $I(1,-l_1,2,-l_2)$ in the two-loop computation. Moreover the analysis of the color factor \eqref{colorNP} shows that the subleading result should have a contribution exactly equal to the leading one. 

Therefore one would be very tempted to trade $I(1,2,-l_1,-l_2)-I(1,2,-l_2,-l_1)$ with\\ $I(1,-l_1,2,-l_2)$. However, equation \eqref{ident} clearly states that these two integrals differ by an imaginary part at order $\e$. Moreover when one tries to use the first combination to compute the subleading part of equation \eqref{FFcut1}, the emerging integral contains spurious propagators.
Explicitly, upon putting back the cut propagators, we can uplift the cut to the two-loop integral
\begin{equation}
\int\frac{d^dl}{(2\pi)^d}\frac{d^dk}{(2\pi)^d}\frac{s\, \Tr(l\,p_1\,p_2) \Tr(k\,p_1\,p_2)}{l^2(l-p_1-p_2)^2(l-p_2)^2(k-l)^2k^2 (k-p_1-p_2)^2(k-p_1)^2}
\end{equation}
looking like the $\textup{\bf DB}_P$ double-box at the special kinematic point $p_3=-p_1$ and $p_4=-p_2$.
Therefore we can derive an expression for such an integral by setting $t=-s$ in the result \eqref{db} for the double-box. This indeed translates into an unphysical imaginary part, which we argue is related to the choice between the two possible integrals with no cut in the $q^2$ channel. Indeed in the kinematic configuration corresponding to the cut\footnote{With our conventions we choose a configuration with $q^2>0$.} the integral $I(1,-l_1,2,-l_2)$ is purely real, as expected since it has no cut in that channel, whereas $I(1,2,-l_1,-l_2)-I(1,2,-l_2,-l_1)$ develops an unphysical imaginary part.

Hence we conclude that a valid prescription to eliminate the ambiguity in the form of the 
one-loop amplitude is to replace the unphysical combination $I(1,2,-l_1,-l_2)-I(1,2,-l_2,-l_1)$ by the integral $I(1,-l_1,2,-l_2)$. This recipe will be further used and developed in the calculation of the amplitude. 

As far as the form factor is concerned, this replacement immediately makes equations \eqref{FFcut1} and \eqref{FFcut2} equal to each other and gives the simple result
\begin{equation}\label{FFcut3}
F^{(2)}(\bar{p}_1,p_2)\Big|_{\textup{$q^2$-cut}} = i\, \frac{\braket{12}\braket{1l_1}}{\braket{2l_1}}\, (N^2-1) \,I(1,-l_1,2,-l_2)
\end{equation}
Then the leading and subleading contributions to the form factor are exactly equal and, using the result of \cite{Brandhuber:2013gda}, we can write down the whole two-loop form factor as
\begin{equation}
F^{(2)}(\bar{p}_1,p_2)=\frac{N^2-1}{k^2}\, \textup{\bf{XT}}(q^2)
\end{equation}
As a consistency check, this integral was also shown in \cite{Brandhuber:2013gda} to be compatible with the vanishing of three-particle cuts.

Substituting the explicit expression of the integral and introducing the same scale $\mu'^{2}= 8 \pi e^{-\gamma_E}\mu^2$ as for the amplitude, the result reads
\begin{equation}
F^{(2)}(\bar{p}_1,p_2)=\frac{N^2-1}{K^2} \left(\frac{-q^2}{{\mu'}^2}\right)^{-2\e} \left[-\frac{1}{(2\e)^2}+\frac32 \log^2 2 +\frac{\pi^2}{6}+{\cal O}(\e)\right]
\end{equation}
This expression agrees with the Feynman diagram computation of \cite{Young:2013hda,Bianchi:2013iha}.

We remark that from this result (and consistently from the leading infrared poles of the amplitude \eqref{eq:result} and of the four-cusps Wilson loop \cite{Bianchi:2013iha}) we can read the leading value of the scaling function $f$ of ABJM at weak coupling.
In contrast to ${\cal N}=4$ SYM in four dimensions\footnote{In ${\cal N}=4$ SYM there are no non-planar corrections to the cusp anomalous dimension up to three loops. At four loops it is not known whether such a contribution could arise. We thank Gang Yang for pointing this out to us.}, a non-planar contribution emerges already at leading order. Defining the 't Hooft coupling $\lambda \equiv N/K$, the scaling function at finite $N$ reads
\begin{equation}
f_{ABJM} = 4 \lambda^2 \left( 1- \frac{1}{N^2} \right) - 24\, \zeta(2)\, \lambda^4 + {\cal O}(\lambda^{6}) + {\cal O}(\lambda^4/N^2)
\end{equation}

\section{Constructing the integrand for the two-loop amplitude}\label{sec:3}

In this section we undertake a constructive approach for the two-loop four-point subleading partial amplitude based on unitarity, in a similar fashion as for the form factor.

Namely we find out what the basis of integrals is for such a quantity, by inspecting its two-particle cuts and uplifting them to proper two-loop integrals. 
This is done by fusing a tree level and a one-loop four-point amplitudes as in Figure \ref{2ptcut_amp}.
We work with superamplitudes on both sides of the cut, integrating over the Grassmann variables of the cut legs. The remaining spinor structure factorizes along with the kinematic pieces of the tree level four-point superamplitude, in front of the integrals.

By combining the different color ordered structures appearing in the complete amplitudes we find the color leading and subleading contributions to a given two-loop partial amplitude. Such a computation demonstrates the absence of double trace contributions at two loops, proportional to $N$. Then both the $N^2$ leading and the $N^0$ subleading single trace contributions are proportional to the same $[1,2,3,4]-[1,4,3,2]$ color structure appearing at tree level. This means that the ratio between the two-loop and the tree level amplitudes can be taken both at the level of the color dressed amplitudes $\tilde{\cal A}$ or the color stripped ones ${\cal A}$. Thus we shall work with partial amplitudes to avoid clutter in the equations. We shall also suppress the coupling constant factor $1/k^2$, which is understood in all two-loop computations.

The procedure outlined above can give an ambiguous answer, which misses integrals that vanish in the selected channel of the cuts. Such ambiguities can be nevertheless fixed by inspection of other cuts, such as three-particle ones, and consistency amongst the various channels.

More severe ambiguities arise from the one-loop function as described in section \ref{sec:1} and can be traced back to the fact that it is subleading in $\epsilon$, whereas we are performing three-dimensional cuts.
Nevertheless we will be able to find a sensible answer by the same prescription used for the form factor, based on the cuts of the one-loop amplitude and the fact that it is expressed in terms of the box function \eqref{eq:boxfunction}.

\subsection{The planar integrand}

\begin{figure}[htbp]
\begin{center}
\includegraphics[width=8. cm]{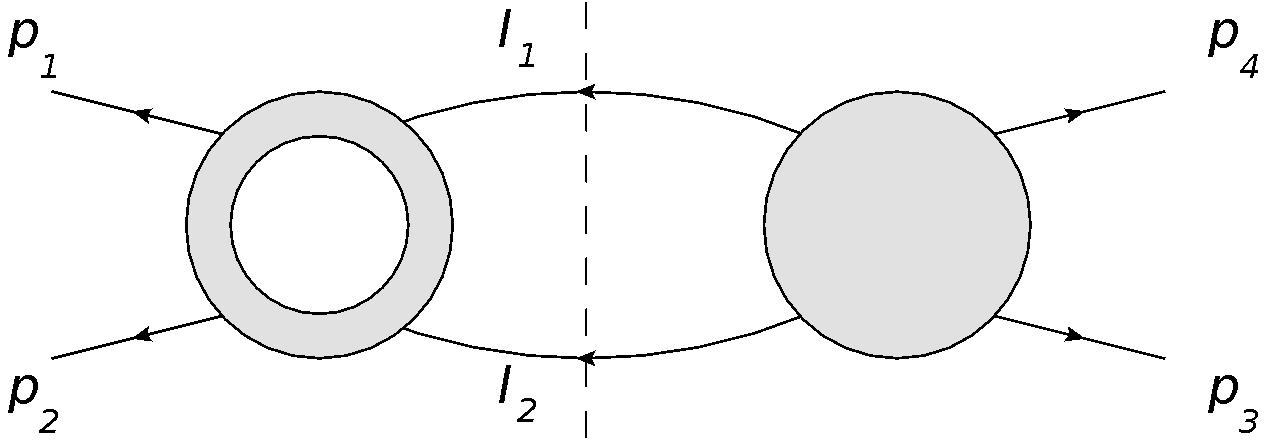}
\caption{One of the two $s$-channel two-particle cuts of the two-loop amplitude. The blob stands for a one-loop amplitude.}
\label{2ptcut_amp}
\end{center}
\end{figure}

As a warm up we first derive the already known planar two-loop contribution \cite{CH,BLMPS1} by fusing the tree level four-point amplitude with the one-loop color leading one.
We spell out the computation for one of the two-particle cuts in the $s$-channel shown in Figure \ref{2ptcut_amp}; all other cuts evaluate similarly upon exchange of momentum labels.

For simplicity we select the color leading contribution (proportional to $N^2$) to the partial amplitude $[1,2,3,4]$
\begin{equation}\label{schleft2}
\left. {\cal A}^{(2)}_4(\bar{1},2,\bar{3},4)\right|_{s-cut}= i\, N^2 \int d^3\eta_{l_1} d^3\eta_{l_2} \,{\cal A}_4^{(0)}(\bar{1},2,-\bar{l_2},-l_1)\, {\cal A}_4^{(0)}(\bar{l_1},l_2,\bar{3},4)\, I(1,2,-l_2,-l_1)
\end{equation}
where $I$ stands for the one-loop box function \eqref{eq:boxfunction}.
Performing the integrals over the spinor variables we get \cite{Brandhuber:2013gda}
\begin{equation}
\int d^3\eta_{l_1} d^3\eta_{l_2}\, {\cal A}_4^{(0)}(\bar{1},2,-\bar{l_2},-l_1)\, {\cal A}_4^{(0)}(\bar{l_1},l_2,\bar{3},4) = {\cal A}_4^{(0)}(\bar{1},2,\bar{3},4)\, \frac{s\, \Tr(l_1\,p_1\,p_4)}{(l_1-p_1)^2(l_1+p_4)^2}
\end{equation}
in the cut configuration and with $l_2=-l_1+p_1+p_2$. Then the planar cut in the $s$-channel reads
\begin{equation}\label{ampsl}
\left. {\cal A}^{(2)}_4(\bar{1},2,\bar{3},4)\right|_{s-cut} = i\, N^2\, {\cal A}_4^{(0)}(\bar{1},2,\bar{3},4)\, \frac{s\, \Tr(l_1\,p_1\,p_4)}{(l_1-p_1)^2(l_1+p_4)^2}\, I(1,2,-l_2,-l_1)
\end{equation}
Using the explicit expression of the one-loop box function\footnote{We cancel the factor $i$ in front of \eqref{ampsl} with that in the integration measure of $I$ in the definition \eqref{eq:boxfunction}, so that the final two-loop integral has the standard measure $\frac{d^dk\, d^dl}{(2\pi)^{2d}}$.} we can manipulate the expression from the cut and, reinstating the cut propagators, we can uplift it to a Feynman integral.
For instance, setting $l\equiv l_1$, we can rewrite the cut integral as
\begin{align}\label{eq:scutplanar1}
& \int \frac{d^dk}{(2\pi)^d}\left[ \frac{s\, \Tr(l\,p_1\,p_4) \left(s\,\Tr(k\,p_1\,p_4)+k^2\, \Tr(p_1\,p_2\,p_4)\right)}{t\, (l+p_4)^2k^2 (k-p_1)^2(k-p_1-p_2)^2(k-l)^2} + 
\frac{s^2}{k^2 (k-p_1-p_2)^2(k-l)^2}\right] =  
\nonumber\\& ~~~~~~~~ = 
\textup{\bf{I}}_P(s,t) \Big|_{l^2=(l+p_{34})^2=0}
\end{align}
which manifestly coincides with the same cut of the integral function $\textup{\bf{I}}_P(s,t) = \textup{\bf DB}_P(s,t) + \textup{\bf DT}_P(s,t)$ appearing in the two-loop amplitude \eqref{eq:planarratio}.
From the form of \eqref{eq:scutplanar1}, uplifting the cut to an integral we only partially reconstruct the correct function, which can be determined imposing consistency with the other two- and three-particles cuts.

Indeed a similar analysis can be carried out for the other two-particle cuts in the $s$- and $t$-channels. Combining the two $s$-channel cuts strongly suggests the final form for the numerator of the double-box appearing in the dual conformally invariant integrand \eqref{eq:dp} of $\textup{\bf I}_P(s,t)$. The cuts in the $t$-channel give rise to  the integral $\textup{\bf I}_P(t,s)$. A nontrivial consistency check that the final combination $\textup{\bf I}_P(s,t)+\textup{\bf I}_P(t,s)$ is the right one is that its three-particle cuts in the $s$- and $t$-channels vanish. This is a consequence of the absence of amplitudes with an odd number of external particles in the ABJM theory. 
Such an analysis was thoroughly carried out in \cite{CH}, which shows that the aforementioned combination of integrals has indeed vanishing three-particle cuts.
We anticipate that the fact that this symmetric combination of double-box integrals is free of three-particle cuts will also provide a useful criterion for fixing their coefficients in the color subleading partial amplitude.

\subsection{The non-planar integrand}

We now turn to the computation of the subleading contributions to the amplitude.
Again, we reconstruct the relevant integral topologies from the analysis of two-particle cuts.

Taking the same channel as in the planar example of the previous section, we can ascertain that only the partial amplitude corresponding to  the $[1,l_1][2,l_2]$ color structure contributes to the subleading part of the amplitude.
Using the prescription explained in section \ref{sec:1} we plug ${\cal A}^{(1)}_4(1,-l_1,2,-l_2)$ in the form \eqref{eq:testamp} into the cut
\begin{align}\label{eq:cutsnp}
\left. {\cal A}^{(2)}_4(\bar{1},2,\bar{3},4)\right|_{s-cut} =& i\, \int d^3\eta_{l_1} d^3\eta_{l_2} {\cal A}^{(0)}_4(\bar{1},2,-\bar{l_2},-l_1) {\cal A}^{(0)}_4(\bar{l_1},l_2,\bar{3},4)\nonumber \\
 &\times (N^2 I(1,2,-l_2,-l_1)+2A(1,-l_1,2,-l_2))
\end{align}
where $A(1,-l_1,2,-l_2)$ was defined in \ref{eq:testamp}. In particular we note that the other subleading partial amplitude $A(1,2,-l_1,-l_2)$ does not contribute to this cut, since combining it with the tree level amplitude gives an identically vanishing double trace color structure $N\,[12][34]$. An analogous contribution from color contracting the tree level amplitude with the leading part of the one-loop amplitude also vanishes by the symmetry properties of the tree level color structures. This implies that no double trace contributions arise from this cut, as anticipated above.

As occurred for the subleading corrections to the Sudakov form factor, we find that the choice $A(1,-l_1,2,-l_2) = I(1,-l_1,-l_2,2) - I(-l_1,2,1,-l_2)$ generates an unphysical result, namely a cut descending from some two-loop integral with a spurious denominator.
We conclude that we have to use the other possible form $A(1,-l_1,2,-l_2) = I(1,-l_1,2,-l_2)$ and hence we consider 
\begin{align}\label{schleft}
\left. {\cal A}^{(2)}_4(\bar{1},2,\bar{3},4)\right|_{s-cut} =&\, i\,\int d^3\eta_{l_1} d^3\eta_{l_2}\, A^{(0)}(\bar{1},2,-\bar{l_2},-l_1)\, A^{(0)}(\bar{l_1},l_2,\bar{3},4)\nonumber \\
 &\times \left(N^2\, I(1,2,-l_2,-l_1)+2\,I(1,-l_1,2,-l_2)\right)
\end{align}
Since we already analyzed the leading part in the previous section, we shall now focus on the subleading contribution only. Performing similar steps as for the color leading case, we find
\begin{equation}\label{ampslnp}
\left. {\cal A}^{(2)}_4(\bar{1},2,\bar{3},4)\right|_{s-cut} = A^{(0)}(\bar{1},2,\bar{3},4)\, {\cal C}_s
\end{equation}
where
\begin{equation}
{\cal C}_s = \frac{2\,i\, s\, \Tr(l_1\,p_1\,p_4)}{(l_1-p_1)^2(l_1+p_4)^2}\, I(1,-l_1,2,-l_2)
\end{equation}
We observe that $I(1,-l_1,2,-l_2)$ is antisymmetric under the exchange of $p_1$ and $p_2$. Consequently we are allowed to take the antisymmetric part of the pre-factor, giving
\begin{equation}\label{eq:sleft}
{\cal C}_s = i\, s\, \left(\frac{ \Tr(l_1\,p_1\,p_4)}{(l_1-p_1)^2(l_1+p_4)^2}-\frac{\Tr(l_1\,p_2\,p_4)}{(l_1-p_2)^2(l_1+p_4)^2}\right)I(1,-l_1,2,-l_2)
\end{equation}
Rewriting it in terms of spinor products and then using the Schouten identity \eqref{eq:schouten}, we find
\begin{equation}
\frac{ \Tr(l_1\,p_1\,p_4)}{(l_1-p_1)^2(l_1+p_4)^2}-\frac{\Tr(l_1\,p_2\,p_4)}{(l_1-p_2)^2(l_1+p_4)^2} = \frac{\braket{12}}{\braket{l_1 1}\braket{l_1 2}}
\end{equation}
After these manipulations $p_4$ disappears completely from the calculation of the cut.
Plugging the explicit form of $I(1,-l_1,2,-l_2)$ (which we shift to $-I(-l_1,2,-l_2,1)$ in order not to have an explicit dependence on $l_2$) we can cast \eqref{eq:sleft} into the form
\begin{equation}
{\cal C}_s = \int \frac{d^d k}{(2\pi)^d}\, \frac{s\, \left(\Tr(p_1\,p_2\,l\,k)+k^2 s \right)}{k^2 (k+l)^2(k+l-p_2)^2(k+p_1)^2}
\end{equation}
Therefore the integral appearing in the $s$-channel cut only depends on $s$. 
If we now reinstate the cut propagators $l^2$ and $(l-p_1-p_2)^2$ (and send $k\rightarrow -k$) we find precisely the same non-planar integral appearing in the computation of the form factor, namely
\begin{equation}
-\int\frac{d^dl}{(2\pi)^d}\frac{d^dk}{(2\pi)^d}\,\frac{s\, \left( \Tr(p_1\,p_2\,l\,k)-s k^2 \right)}{l^2(l-p_{12})^2k^2(l-k)^2(k-l+p_2)^2(k-p_1)^2} = -\textup{\bf{XT}}(s)
\end{equation}
A completely parallel analysis can be performed for the other $s$- and $t$-channel two-particle cuts, leading again to integrals of the topology $\textup{\bf{XT}}$ in all the four cyclic permutations of the external legs. Since such integrals depend on one invariant only these are automatically pairwise identical and the result can be summarized as $-2\,\textup{\bf{XT}}(s)-2\,\textup{\bf{XT}}(t)$.

The two-particle cuts in the $u$-channel are slightly different from the others. From the analysis of the color algebra it turns out that there is no leading contribution, unlike the $s$- and $t$-channel cases. Explicitly, we can select one of the $u$-channel cuts and its expression reads
\begin{align}
 \left. {\cal A}^{(2)}_4(\bar{1},2,\bar{3},4)\right|_{u-cut} =& \, i\, \int d^3\eta_{l_1} d^3\eta_{l_2}\, {\cal A}^{(0)}_4(\bar{1},-l_1,\bar{3},-l_2)\, {\cal A}^{(0)}_4(\bar{l_1},2,\bar{l_2},4)\nonumber \\
 & \times \left( 2\, A(1,3,-l_2,-l_1) + 2\, A(1,3,-l_1,-l_2) \right)
\end{align}
In this case one can show that
\begin{multline}
\int d^3\eta_{l_1} d^3\eta_{l_2}\, {\cal A}^{(0)}_4(\bar{1},-l_1,\bar{3},-l_2)\, {\cal A}^{(0)}_4(\bar{l_1},2,\bar{l_2},4)\\
=\, {\cal A}^{(0)}_4(\bar{1},2,\bar{3},4)\, \frac12\, \left( \frac{u\, \Tr(l_1\,p_3\,p_4)}{(l_1-p_3)^2(l_1+p_4)^2}+\frac{u\, \Tr(l_2\,p_3\,p_4)}{(l_2-p_3)^2(l_2+p_4)^2}\right)
\end{multline}
where the factor $\frac12$ has to be introduced to account for identical particles running in the loop \cite{Brandhuber:2013gda}.
With this prescription we find
\begin{equation}
\left. {\cal A}^{(2)}_4(\bar{1},2,\bar{3},4)\right|_{u-cut} = {\cal A}^{(0)}_4(\bar{1},2,\bar{3},4)\, {\cal C}_u
\end{equation}
where
\begin{equation}
{\cal C}_u = i\, \left(\frac{u\, \Tr(l_1\,p_3\,p_4)}{(l_1-p_3)^2(l_1+p_4)^2}+(l_1\leftrightarrow l_2)\right) 
\times \Big( A(1,3,-l_2,-l_1) + (l_1\leftrightarrow l_2) \Big)
\end{equation}
In both the pre-factor and in the combination of the $A$ functions it can be proved that the manifest symmetry under the exchange of $l_1$ and $l_2$ is equivalent to that under $p_1 \leftrightarrow p_3$, giving explicitly
\begin{equation}
{\cal C}_u = i\, \left(\frac{u\, \Tr(l_1\,p_3\,p_4)}{(l_1-p_3)^2(l_1+p_4)^2} + \frac{u\, \Tr(l_1\,p_1\,p_4)}{(l_1-p_1)^2(l_1+p_4)^2} \right) \, \left[ A(1,3,-l_2,-l_1) + A(3,1,-l_2,-l_1) \right]
\end{equation}
Putting a common denominator in the pre-factor, one obtains a sum of spinor products in the numerator, which can be further massaged by means of the Schouten identity \eqref{eq:schouten}.
We can separate two different pieces multiplying $A(1,3,-l_2,-l_1)$ and $A(3,1,-l_2,-l_1)$ respectively and act with the Schouten identity in a different fashion in each of the parts.  
In the end re-combining everything we obtain the following expression
\begin{align}
{\cal C}_u = &\,\, 2\,i\, \left[ \frac{u\, \Tr(l_1\,p_3\,p_4)}{(l_1-p_3)^2(l_1+p_4)^2}\, A(3,1,-l_2,-l_1) + (p_1 \leftrightarrow p_3 ) \right]\nonumber \\
& + \frac{u\, \Tr(l_1\,p_1\,p_3)}{(l_1-p_1)^2(l_1-p_3)^2}\, \left[ A(1,3,-l_2,-l_1) - A(3,1,-l_2,-l_1) \right]
\end{align}
By means of the symmetry and cut properties of $A$, the combination in the second line is actually equivalent to $A(-l_1,3,-l_2,1)$. 
At this point we replace the objects $A$ with corresponding box integrals $I$. We find that the only sensible choice, which does not produce any unphysical integrals, is $A(3,1,-l_2,-l_1)\rightarrow I(3,1,-l_2,-l_1)$ and $A(-l_1,3,-l_2,1)\rightarrow I(-l_1,3,-l_2,1)$. 
After analogous cosmetics as for the other channels, we arrive at the following form for the cut
\begin{align}
{\cal C}_u = &\,\, \int \frac{d^d k}{(2\pi)^d}\, \Bigg\{ 2\, u \left(\frac{\Tr(l\,p_3\,p_4) \left[ u\, \Tr(k\,p_3\,p_4)+ k^2\Tr(p_3\,p_1\,p_4)\right]}{s\, (l+p_4)^2k^2 (k-p_3)^2(k-p_1-p_3)^2(k-l)^2} + (p_1 \leftrightarrow p_3)\right) + \non \\& + 4\, \frac{u^2}{(k-l)^2 k^2 (k-p_1-p_3)^2} + \frac{u\, (\Tr(p_1\,p_3\,l\,k) + u\, k^2)}{k^2 (k+l)^2(k+l-p_3)^2(k+p_1)^2} \Bigg\}
\end{align}
The part of the cut in the first line, along with the first contribution of the second one, clearly arises from the planar double-box topology $2\, \textup{\bf I}_P(u,s)+2\, \textup{\bf I}_P(u,t)$, as can be seen by comparing to \eqref{eq:scutplanar1} and making suitable replacements of momentum labels.
The remaining part of the cut uplifts to the integral $-\textup{\bf XT}(u)$.

By inspection of the other $u$-channel cut we find again the same planar double-box topologies and another $-\textup{\bf XT}(u)$, which however originates from a different choice of external legs and consequently has to be counted twice.

Combining everything gives
\begin{equation}
2 \left( \textup{\bf I}_P(u,s) + \textup{\bf I}_P(u,t) - \textup{\bf XT}(s) - \textup{\bf XT}(t) - \textup{\bf XT}(u)\right)
\end{equation}
It remains to check the consistency of this combination of integrals by imposing the vanishing of three-particle cuts. This is automatic for the $\textup{\bf XT}$ integrals, as verified in \cite{Brandhuber:2013gda} because they appear in the computation of the form factor.
The sum $\textup{\bf I}_P(u,s) + \textup{\bf I}_P(u,t)$ has non-vanishing three-particle cuts. The cut analysis performed for the leading amplitude then suggests how to cancel them, namely we add to them their counterpart with symmetrized invariants
\begin{equation}
2 \left( \textup{\bf I}_P(u,s) + \textup{\bf I}_P(s,u) + \textup{\bf I}_P(u,t) + \textup{\bf I}_P(t,u) \right)
\end{equation}
This can not be the complete answer since the addition of e.g.~the $\textup{\bf I}_P(s,u)$ integral should have been detected by the $s$-channel quadruple cut, and similarly for $\textup{\bf I}_P(t,u)$. We can solve this puzzle by adding a contribution which preserves the vanishing of three-particle cuts and cancels the contribution of $\textup{\bf I}_P(s,u)$ and $\textup{\bf I}_P(t,u)$ to the quadruple cuts. This contribution, as was shown in \cite{CH} is given by $-2\, \textup{\bf I}_P(s,t) -2\, \textup{\bf I}_P(t,s)$. 
Indeed the difference $\textup{\bf I}_P(s,u)-\textup{\bf I}_P(s,t)$ possesses a vanishing quadruple cut, due to its symmetries.
Henceforth our final result for the leading and subleading contribution to the two-loop amplitude reads 
\begin{align}\label{eq:result2}
{\cal M}_4 = & \, 
N^2 \left( \textup{\bf I}_P(s,t) + \textup{\bf I}_P(t,s) \right) -2 \left( \textup{\bf XT}(s) + \textup{\bf XT}(t) + \textup{\bf XT}(u) \right) + \nonumber\\&
 + 2 \left( \textup{\bf I}_P(u,s) + \textup{\bf I}_P(s,u) + \textup{\bf I}_P(u,t) + \textup{\bf I}_P(t,u) - \textup{\bf I}_P(s,t) - \textup{\bf I}_P(t,s) \right)
\end{align}
We can successfully check that this combination indeed reproduces the known result from a Feynman diagram computation \eqref{eq:result}, which provides the best test on the validity of our procedure.

In particular we note that, as in the Feynman diagram computation, the non-planar topologies only contribute through the simple integrals $\textup{\bf XT}$, depending on one scale only. 

Finally, setting $N=2$ in \eqref{eq:result2}, one can get the BLG two-loop four-point amplitude ratio
\begin{align}
\mathcal{M}^{BLG}_4=2\, \Big(&\textup{\bf{I}}_P(u,s)+\textup{\bf{I}}_P(u,t)+\textup{\bf{I}}_P(s,u)
+\textup{\bf{I}}_P(s,t)+\textup{\bf{I}}_P(t,u)+\textup{\bf{I}}_P(t,s)\non \\
&-\textup{\bf{XT}}(s)-\textup{\bf{XT}}(t)-\textup{\bf{XT}}(u)\Big)
\end{align}
which takes a manifestly totally symmetric form at the level of the integrals.

\section{An alternative ansatz for the integral basis}\label{sec:4}

In this section we re-derive the result for the two-loop subleading partial amplitude following a reverse logic with respect to the previous one. Namely, we formulate a guess on the integral basis for the amplitude and fix the relative coefficients by demanding that the quadruple and triple cuts are satisfied.

Given that the color leading amplitude can be expressed in terms of a double-box with a particular numerator making it dual conformally invariant, the natural guess is to expect a non-planar version of it to appear in the subleading contribution.
It remains to determine its correct numerator.
Unfortunately, unlike the planar case, we cannot use dual conformal invariance as a guiding principle, since we do not expect the non-planar correction to enjoy this symmetry, by analogy with the four-dimensional case.

For Yang-Mills theory the BCJ identities can be used to find relations among the numerators in the planar and non-planar integrals \cite{Bern:2008qj}.
The BCJ relations connect different tree level partial amplitudes through a Jacobi identity. By means of unitarity one can divide a loop amplitude into tree level sub-amplitudes and apply the BCJ on these.
This in turn imposes constraints on the numerators of triples of loop integrals.

Analogous relations between color ordered partial amplitudes have been derived for ABJM \cite{Bargheer:2012cp} and more general models with  bi-fundamental scattered particles \cite{Huang:2013kca}. In contrast to Yang-Mills theory, which possesses a Lie algebra color structure, bi-fundamental theories like ABJM have an underpinning three-algebra pattern, with a four-indexed structure constant.
This entails BCJ identities that are present for the ABJM theory up to six points and extend to all multiplicities for BLG theory.

The nontrivial six-point identity involves four terms and is diagrammatically formulated in terms of four-line vertices. These are absent in our integral topology, therefore we can't apply them.

\begin{figure}[htbp]
\begin{center}
\includegraphics[width=11. cm]{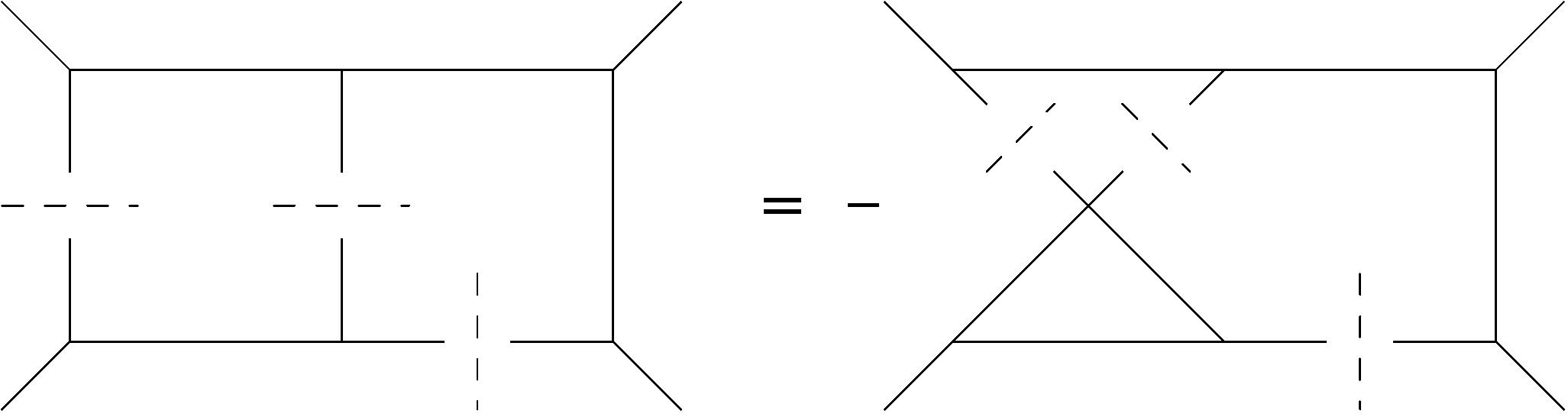}
\caption{The numerators of planar and non-planar topologies are constrained from color-kinematics duality.}
\label{subamplitude}
\end{center}
\end{figure}

More simply we take inspiration from the BLG theory where the tree level four-point amplitude can be expressed as depending on a totally antisymmetric structure constant $f^{abcd} \propto \epsilon^{abcd}$.
We consider the planar integral $\textup{\bf DB}_P$ whose explicit form we recall here for convenience:
\begin{equation}\label{eq:dbnp}
\textup{\bf{DB}}_P(s,t)=\int\frac{d^dl}{(2\pi)^d}\frac{d^dk}{(2\pi)^d}\frac{\left[s\, \Tr(l\,p_1\,p_4)+l^2\Tr(p_1\,p_2\,p_4)\right]\left[s\, \Tr(k\,p_1\,p_4)+k^2 \Tr(p_1\,p_2\,p_4)\right]}{t\,l^2(l+p_3+p_4)^2(l+p_4)^2(k-l)^2k^2 (k-p_1-p_2)^2(k-p_1)^2}
\end{equation}
Next we can perform a cut in the amplitude isolating a four-point sub-amplitude, as in Figure \ref{subamplitude}. For BLG theory such an amplitude is totally antisymmetric under the exchange of external labels. We can now obtain a non-planar integral topology by permuting two legs of this amplitude as shown in Figure \ref{subamplitude}.
This involves the replacement of the propagator $(k-p_{12})^{-2}$ by  $(k-l-p_2)^{-2}$, which can also be seen as replacing $p_2 \rightarrow -p_{12}+l$ in the cut sub-amplitude. 

Then, using the antisymmetry of the four-point sub-amplitude, we see that in BLG theory an identity should hold between the numerators of the planar and non-planar integrals.
Namely the two numerators should be opposite, provided the cut conditions hold and the aforementioned replacement in the involved momenta is made, so that the cut momenta coincide. 
This constraint can be surely fulfilled (up to a sign) if we just perform the replacement $p_2 \rightarrow -p_{12}+l$ in the relevant part of the planar integral numerator. As a result we obtain the non-planar double-box integral
\begin{equation}
 \textup{\bf{DB}}_{NP} \equiv \int\frac{d^dl\,d^dk}{(2\pi)^{2d}}\frac{[s\, \Tr(l\,p_1p_4)+l^2\Tr(p_1\,p_2\,p_4)] [(l-p_2)^2 \Tr(k\,p_1p_4)+ k^2\Tr(p_1(l-p_2)\,p_4)]}{t\, l^2(l+p_3+p_4)^2(l+p_4)^2k^2 (k-l)^2(k-l+p_2)^2(k-p_1)^2}
\end{equation}
As in the planar case this integral could be accompanied by a simpler topology.
Again we make an ansatz for it by performing the same operation as above on the integral $\textup{\bf DT}_P$, giving
\begin{equation}
\textup{\bf{DT}}_{NP}(s) \equiv \int\frac{d^dl}{(2\pi)^d}\frac{d^dk}{(2\pi)^d}\frac{s\, (l-p_2)^2}{l^2(l+p_3+p_4)^2(k-l)^2 k^2 (k-l+p_2)^2}
\end{equation}
According to the BLG BCJ identity above, the non-planar integrals should appear with a relative minus sign with respect to the planar ones.
Since we also want to extend the amplitude computation to the ABJM theory, we disregard such signs and simply propose that a combination of the planar $\textup{\bf DB}_P$, $\textup{\bf DT}_P$ and the non-planar $\textup{\bf DB}_{NP}$ and $\textup{\bf DT}_{NP}$ integrals\footnote{The topology of $\textup{\bf DT}_{NP}$ is actually planar and we have loosely referred to it as non-planar in the sense that it emerges from the same operation transforming the planar double-box into a non-planar one.} gives the subleading partial amplitude at two loops.

\begin{figure}[htbp]
\begin{center}
\includegraphics[width=14. cm]{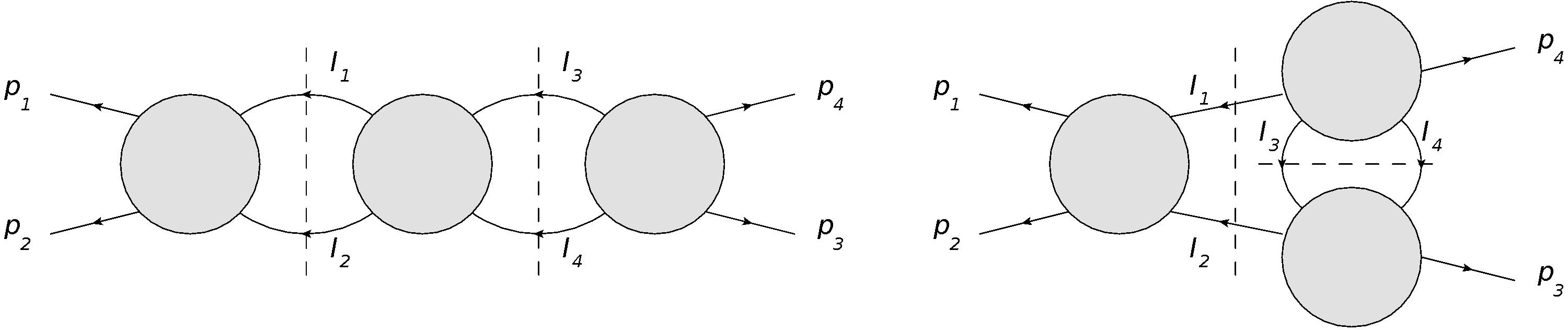}
\caption{Quadruple cuts of the two-loop amplitude. The first one receives contributions from the planar integrals only, the second from both planar and non-planar.}
\label{doublecut_amp}
\end{center}
\end{figure}

In fact, by performing quadruple and triple cuts as in Figure \ref{doublecut_amp}, we are able to fix the relevant coefficients of the integrals. The requirement is that the former give the product of three four-point tree level amplitudes, whereas the latter vanish. This must be the case since three-particle cuts separate the two-loop amplitude into two five-point tree level amplitudes, which identically vanish in ABJM \cite{BLM,CH}.
The result of such an analysis reads:
\begin{align}\label{eq:result3}
\mathcal{M}_4 =& N^2\Big(\textup{\bf{I}}_P(s,t)+\textup{\bf{I}}_P(t,s)\Big) - 4\, \Big(\textup{\bf{I}}_{NP}(s,t,u)+\textup{\bf{I}}_{NP}(u,s,t)+\textup{\bf{I}}_{NP}(t,u,s)\Big) + \nonumber\\& + 2\, \Big(\textup{\bf{I}}_P(u,s)+\textup{\bf{I}}_P(u,t)+\textup{\bf{I}}_P(s,u)
-\textup{\bf{I}}_P(s,t)+\textup{\bf{I}}_P(t,u)-\textup{\bf{I}}_P(t,s)\Big)
\end{align}
where we have defined
\begin{equation}
 \textup{\bf{I}}_{NP}(s,t,u) \equiv \textup{\bf{DB}}_{NP}(s,t,u) + \textup{\bf{DT}}_{NP}(s)
\end{equation}
in a similar fashion to the planar case.
The sum over the $S_3$ permutations of the external legs in the $\textup{\bf{I}}_{NP}$ integrals has been rewritten in \eqref{eq:result3} as a sum over the cyclic ones, adding a factor of 2, thanks to their symmetry properties.

We now check that this is indeed a correct integral representation of the amplitude by explicitly computing the two-loop integrals and matching to \eqref{eq:result}.
The hardest task is to evaluate the two-loop non-planar double-box, which displays a nasty numerator.
We deal with it by first expanding the product of the traces into scalar products, and then we turn them into inverse propagators. Many of them cancel against some denominator producing simpler topologies, but we are also left with integrals with up to three irreducible numerators.
We simplify this massive combination of scalar integrals by reducing them to master integrals via integration by parts identities. We employ the \texttt{FIRE} package  \cite{Smirnov:2008iw} to carry out such a program, whose details can be found in Appendix \ref{app:2}. Then the problem boils down to the computation of a restricted set of three-dimensional master integrals, which we perform in Appendix \ref{app:3}, by writing their Mellin-Barnes representations and solving these integrals by repeated use of the Barnes lemmas and their corollaries.
The results for the non-planar double-box integral $\textup{\bf DB}_{NP}$ and for $\textup{\bf DT}_{NP}$ turn out to be rather simple and are explicitly given in \eqref{dbnp} and \eqref{dtnp}.

In particular, we observe that the combination $\textup{\bf DB}_{NP}+\textup{\bf DT}_{NP}$ appearing in the amplitude dramatically simplifies giving (up to subleading in $\epsilon$ terms)
\begin{equation}\label{inp}
\textup{\bf{I}}_{NP}(s,t,u) = -\frac{1}{16\pi^2}\left(\frac{-s e^{\g_E}}{4\pi\mu^2}\right)^{-2\e}\left(\frac{1}{8\e^2}+\frac{2\log2+\log\frac{u}{t}}{8\e}-\frac12\z_2 -\frac12\log^2 2 +\frac12\log\frac{s}{u} \right)
\end{equation}
We further observe that the sum over all permutations of external legs of such objects reduces to
\begin{equation}\label{eq:integralidentity}
\sum_{\s\in S_3}\textup{\bf{I}}_{NP}(\s(s,t,u)) = \textup{\bf{XT}}(s)+\textup{\bf{XT}}(t)+\textup{\bf{XT}}(u)
\end{equation}
which is manifestly maximally transcendental.
Given the identity \eqref{eq:integralidentity} it is then immediate to map the result of the amplitude in terms of double-boxes (\ref{eq:result3}) into the previous expression (\ref{eq:result2}), thus establishing their equivalence.
We have therefore verified that our ansatz for the integral basis of the subleading partial amplitude is indeed correct, and provided an alternative formulation for it.

Finally we comment on the BLG amplitude. Again we can obtain it by setting $N=2$ in the ABJM result, giving
\begin{equation}\label{eq:result3BLG}
\mathcal{M}_4 = 2\Big(\textup{\bf{I}}_P(s,t)-\textup{\bf{I}}_{NP}(s,t,u)\Big) + \mathrm{perms}(2,3,4)
\end{equation}
In such a form it is clear that the numerator identity depicted in Figure \ref{subamplitude} between the planar and non-planar integrals is verified.
Indeed, by cutting the planar double-box integral $\textup{\bf I}_P$ we can isolate the four-point sub-amplitude of Figure \ref{subamplitude}. In BLG theory it is proportional to the totally antisymmetric three-algebra structure constant $f^{abcd}$. Obtaining the non-planar topology corresponds to crossing two loop momenta of such  a four-point sub-amplitude, which entails a minus sign change in its color factor $c_i \rightarrow -c_i$.
Color-kinematics duality then prescribes that the same change should be paralleled by the numerators, namely $N_i \rightarrow -N_i$, once loop momenta have been identified.
We observe that the relative minus sign between the integral topologies $\textup{\bf I}_P$ and $\textup{\bf I}_{NP}$ respects this principle.
Therefore we propose that this is the correct BCJ form of the two-loop amplitude ratio.
It would be interesting to re-derive this result from unitarity within the $SO(4)$ description of BLG theory (instead of the ABJM or bi-fundamental formalism we adopted here) where the three-algebra color structure is more manifest.

Believing that a double copy of BLG amplitudes can reproduce gravity ones also at loop level, one should then recover the two-loop four-point amplitude of ${\cal N}=16$ three-dimensional supergravity by dropping color factors and properly squaring the numerators.
The resulting integrals would display quite complicated numerators and we shall not attempt to undertake such a program in this paper.

\section{The ${\cal N}=8$ SYM complete amplitude and its double copy}\label{sec:5}

${\cal N}=8$ SYM can be obtained by dimensional reduction of ${\cal N}=4$ SYM compactifying one dimension. Compactifying time gives an Euclidean three-dimensional model, which can be interpreted as the infinite temperature limit of ${\cal N}=4$ SYM. Here we shall compactify a spatial direction and consider Lorentzian ${\cal N}=8$ SYM.
As pointed out in \cite{Lipstein:2012kd} such a reduction does not affect the Feynamn rules. Therefore one can argue that the amplitudes in the two theories can be expressed by the same diagrams upon restricting the kinematics and loop momenta to three dimensions.
In particular, we can borrow results from ${\cal N}=4$ SYM and compute loop amplitudes for ${\cal N}=8$ SYM, by solving the relevant integrals in three dimensions.
This entails that, in the planar limit, ${\cal N}=8$ SYM integrals are dual conformally covariant, though not invariant.
At one loop, the computation of boxes in three dimensions shows that amplitudes are finite and the MHV ones (including the four-point case) are subleading in the dimensional regularization parameter $\epsilon$ \cite{Lipstein:2012kd}.
The computation of the four-point two-loop amplitude was performed in \cite{Bianchi:2012ez}, in the planar limit, where the only integral appearing in the computation was a scalar planar double-box. This integral was solved in three dimensions by Mellin-Barnes techniques \cite{Bianchi:2012ez} and is a master integral for the reduction of the integral $\textup{\bf DB}_P$ governing the ABJM amplitude in the large $N$ limit. 

If we now turn to color subleading contributions, the non-planar double-box topology appears.
Indeed, following dimensional reduction from $\mathcal{N}=4$ SYM, we can expand the complete four-point amplitude in color space. Taking for instance the notation of \cite{Naculich:2008ys} and the results reported therein, the single trace subleading amplitude ${\cal A}^{(2,2)}_{[1]}$ reads
\begin{align}
{\cal A}^{(2,2)}_{[1]} = & -\frac12 s\, t\, {\cal A}^{(0)}_{[1]} \left[
s\left( \textup{\bf LAD}(s,t) + \textup{\bf NPL}(s,t) + \textup{\bf LAD}(s,u) + \textup{\bf NPL}(s,u) \right) 
\right. \nonumber\\ &
+ t\left( \textup{\bf LAD}(t,s) + \textup{\bf NPL}(t,s) + \textup{\bf LAD}(t,u) + \textup{\bf NPL}(t,u) \right)
\nonumber\\& \left.
-2\, u\left( \textup{\bf LAD}(u,s) + \textup{\bf NPL}(u,s) + \textup{\bf LAD}(u,t) + \textup{\bf NPL}(u,t) \right) \right]
\end{align}
and the double trace contribution ${\cal A}^{(2,1)}_{[7]}$
\begin{align}
{\cal A}^{(2,1)}_{[7]} = & -2\,i\,K \left[
s\left( 3\, \textup{\bf LAD}(s,t) + 2\, \textup{\bf NPL}(s,t) + 3\, \textup{\bf LAD}(s,u) + 2\, \textup{\bf NPL}(s,u) \right) 
\right. \nonumber\\& \left.
- t\left( \textup{\bf NPL}(t,s) + \textup{\bf NPL}(t,u) \right)
-u\left( \textup{\bf NPL}(u,s) + \textup{\bf NPL}(u,t) \right) \right]
\end{align}
where $\textup{\bf LAD}$ and $\textup{\bf NPL}$ stand for the planar and non-planar ladder master integrals.
Their solutions are reported in \eqref{eq:ladder} and \eqref{eq:npladder}, respectively.
The combinations appearing in the partial amplitudes give rise to complicated expressions, which are not particularly illuminating.
Contrary to the ${\cal N}=4$ SYM, they are not maximally transcendental, but this does not come as a surprise, since the color leading amplitude of ${\cal N}=8$ SYM in three dimensions does not show uniform transcendentality either.

We observe a cancellation of the double $\epsilon$ poles in the infrared divergent part of the subleading amplitude, which possesses a milder $1/\epsilon$ behaviour.
This strictly resembles an analogous phenomenon of ${\cal N}=4$ SYM amplitudes at one loop. In particular it shows that the cusp anomalous dimension of ${\cal N}=8$ SYM in three dimensions does not receive nonplanar contributions at leading order.

The softer infrared behaviour should also occur for the four-point two-loop amplitude of ${\cal N}=16$ supergravity \cite{Marcus:1983hb} in three dimensions, to which the gauge theory amplitude is meant to be tightly connected by a squaring procedure, once it is expressed in a color-kinematics dual fashion. 

For the complete SYM amplitude such a form is obtained by taking the combination of the two integral topologies, dressing them with a color factor by associating a structure constant $f^{abc}$ to each three-point vertex of the diagram, and summing over the permutations of external legs. At two loops and four points this takes the form \cite{Bern:1998ug}
\begin{equation}
\tilde{\cal A}_4^{(2)} = s\, t\, {\cal A}_4^{(0)}\, \left[ c^P_{1234}\, s\, \textup{\bf LAD}(s,t) + c^{NP}_{1234}\, s\, \textup{\bf NPL}(s,t) + \mathrm{perm}(2,3,4) \right]
\end{equation}  
where $c^P$ and $c^{NP}$ stand for the color factors of the planar and non-planar ladders respectively, and a sum over the permutations of the last three external legs is performed.
By replacing the color factors with another power of the numerator appearing in the integral, we obtain the ${\cal N}=16$ supergravity four-point amplitude \cite{Bern:1998ug}
\begin{equation}
M_4^{(2)} = - s\, t\, u\, M_4^{(0)}\, \left[ s^2\, \textup{\bf LAD}(s,t) + s^2\, \textup{\bf NPL}(s,t) + \mathrm{perm}(2,3,4) \right]
\end{equation}
By explicitly plugging the expression for the planar and non-planar double-box integrals we find
\begin{align}
\frac{M_4^{(2)}}{M_4^{(0)}} &= \frac{1}{16\pi^2}\left(\frac{-s}{\mu'^2}\right)^{-2\epsilon} \left[-\frac{s}{\epsilon ^2}-\frac{9\, s}{2\epsilon } + \frac{\pi ^2}{6} \frac{(s^2+t^2+u^2)^2}{s\,t\,u} - \frac {3 s^2+3 t^2-7 u^2}{2\,u} \log ^2\left(\frac{s}{t}\right)\right] \nonumber\\
&+ \mathrm{cycl}(2,3,4)
\end{align}
where the sum is performed over the two other cyclic permutations of the last three external momenta and $\mu'$ is the same redefinition of the dimensional regularization mass scale as for the ABJM amplitude \eqref{eq:planarresult}.
We observe that the leading infrared divergence vanishes and only single poles in $\epsilon$ are present.
Moreover the result is almost uniformly transcendental, except for the $1/\epsilon$ pole.

\section{Conclusions}

In ABJM theory we have computed the color subleading two-loop corrections to the Sudakov form factor and the four-point amplitude via unitarity.
We pointed out that applying two-particle cuts in strictly three dimensions leads to ambiguities due to the one-loop four-point amplitude being subleading in the dimensional regularization parameter. This is a peculiar situation of these three-dimensional models, which does not occur in four dimensions.
We proposed a prescription to overcome these difficulties without resorting to $d$-dimensional unitarity and applied it to reconstruct the integral basis involved in the non-planar contributions to the form factor and the amplitude. By setting the rank of the gauge group to 2, we also provided an expression for the four-point amplitude in BLG theory.
As a check, we verified that our final results coincide with previous computations using Feynman diagrams.

For the four-point amplitude, another path to circumvent the ambiguities due to the one-loop amplitude consists of performing quadruple cuts, isolating tree level sub-amplitudes only.
From these it is rather difficult to reconstruct the numerators of the integrals to which they are uplifted when reintroducing the cut propagators. Given that we already know the integral basis of the planar part of the amplitude, we formulated an ansatz for the non-planar integrals, inspired by color-kinematics duality.
Then we fixed the coefficients of these integrals by generalized unitarity and successfully tested the correctness of such a combination against the known results.
This involved solving a non-planar double-box with a complicated numerator, which we treated via reduction to master integrals. In turn, we computed the latter using Mellin-Barnes techniques.

For the BLG amplitude we proposed that the integral representation we pointed out obeys color-kinematics duality at loop level, which is based on the three-algebra structure underlying the BLG model.

Finally, we exploited our results for three-dimensional integrals to compute the color subleading contributions to the two-loop four-point amplitude in ${\cal N}=8$ SYM. 
We wrote the complete amplitude in a fashion respecting color-kinematics duality,
now in the traditional Lie algebra environment, and squared the numerators of its integrals to present an explicit expression for the two-loop four-point amplitude in ${\cal N}=16$ supergravity.

\bigskip
In \cite{Huang:2012wr} it was pointed out that BLG tree level amplitudes also reproduce those of ${\cal N}=16$ supergravity by a double copy, given the uniqueness of this theory. In this case the starting point is a form possessing color-kinematics duality with respect to a three-algebra structure \cite{Bargheer:2012gv,Huang:2013kca}.
Unitarity implies this to propagate to loop level also, therefore it would be interesting to check whether the supergravity two-loop amplitudes obtained from squaring BLG and ${\cal N}=8$ SYM are indeed identical.
In the SYM case such a squaring does not change the relevant integrals since at two-loops only numerators made of invariants of the external momenta appear.
On the contrary, for the BLG theory this procedure would probably lead to new integrals with complicated numerators.
Their solution goes beyond the aim of this paper and we leave it for future research.

Gravity theories in three dimensions are also power counting non-renormalizable as in four. Therefore the study of their ultraviolet behaviour is intriguing. In particular the existence of two different double copy formalisms is a distinctive feature of three dimensions and its consequences on the the ultraviolet properties of three-dimensional supergravity are worth investigating.

\section*{Acknowledgements}

We thank Gang Yang, Valentina Forini, Ben Hoare, Matias Leoni, Lorenzo Magnea, Marco Meineri, Silvia Penati, Jan Plefka and Gabriele Travaglini for very useful discussions. The work of LB is funded by DFG via the Emmy Noether Program ``Gauge Fields from Strings''. 
The work of MB has been supported by the Volkswagen-Foundation.

\vfill
\newpage

\appendix

\section{Notation and conventions}\label{app:1}

We work with the Minkowski metric $g_{\mu\nu}={\rm diag}\{1,-1,-1\}$.
and the totally antisymmetric tensor $\varepsilon^{\mu\nu\rho}$, defined by
$\varepsilon_{012}=\varepsilon^{012}=1$. Spinor indices are raised and lowered as $\l_{\alpha}=\varepsilon_{\alpha \beta} \l^{\beta}$ with $\varepsilon_{12}=\varepsilon^{12}=1$.

On-shell solutions of the fermionic equations of motion are expressed in terms of $SL(2,\mathbb{R})$ commuting spinors $\l_\a$. The same quantities allow one to write on-shell momenta as 
\begin{equation}
p_{\a\b}=(\gamma^\mu)_{\a\b}\ p_\mu
\end{equation}
where the set of $2 \times 2$ gamma matrices are chosen to satisfy 
\begin{equation}
\left(\g^{\mu}\right)^{\a}_{~\g}\, \left(\g^{\nu}\right)^{\g}_{~\b} = -g^{\m\n}\, \d^{\a}_{~\b} - \e^{\m\n\rho}\, \left(\g_{\rho}\right)^{\a}_{~\b}
\end{equation}
An explicit set of matrices is $( \gamma^{\mu} )_{\a\b}= \{ \s^0, \s^1, \s^3 \}$. 

We define spinor contractions as
\begin{equation}
\langle i \, j\rangle=-\langle j \, i\rangle \equiv \lambda^{\alpha}_i\lambda_{\alpha j}=\epsilon_{\alpha\beta}
\lambda^{\alpha}_i \lambda^{\beta}_j
\end{equation}
They obey the Schouten identity
\begin{equation}\label{eq:schouten}
\braket{a b} \braket{c d} + \braket{a c} \braket{d b} + \braket{a d} \braket{b c} = 0
\end{equation}
Thus for any pair of on-shell momenta we write
\begin{equation}
p_{ij}^2 \equiv (p_i+p_j)^2 = 2 \, p_i \cdot p_j =  p_i^{\a\b} \, (p_j)_{\a\b} = -\langle i \, j\rangle^2
\end{equation}
For positive energy spinors are real, whereas for negative energy they are imaginary.

Traces:
\begin{equation}
\braket{ij} \braket{ji}\ =\ -2\ p_i \cdot p_j
\end{equation}
\begin{equation}
\braket{ij} \braket{jk} \braket{ki}\ =\ \Tr (p_i\ p_j\ p_k) = 2\, \epsilon(i,j,k)
\end{equation}
\begin{align}
& \braket{ij} \braket{jk} \braket{kl} \braket{li}\ =\ \Tr (p_i\ p_j\ p_k\ p_l) = 
\nonumber\\&
2\, \left[
\sp{p_i}{p_j}\sp{p_k}{p_l} + \sp{p_i}{p_l}\sp{p_j}{p_k} - \sp{p_i}{p_k}\sp{p_j}{p_l}
\right]
\end{align}

For definiteness we will choose a regime where
\begin{equation}
\braket{12} = \braket{43} \quad \braket{23} = \braket{41} \quad \braket{13} = \braket{24} 
\end{equation}

We will use the four-point superamplitude
\begin{equation}
{\cal A}_4 = i\, \frac{\delta^{(3)}(P)\delta^{(6)}(Q)}{\braket{12}\braket{23}}
\end{equation}

At loop level our integrals are normalized with the measure
\begin{equation}
\int \frac{d^{3-2\epsilon}k}{(2\pi)^{3-2\epsilon}}
\end{equation}
for each loop integration.

\section{Reduction to master integrals}\label{app:2}

The unitarity based computation of the two-loop contributions to the Sudakov form factor and four-point amplitude in ABJM produces integrals with tensor structure.
We deal with them by reduction to master integrals via integration by parts identities.
We use the package \texttt{FIRE} \cite{Smirnov:2008iw} to automatically perform this task. We list here the integrals we had to reduce for our computation:
\begin{align}
\textup{\bf{DB}}_P(s,t)&=\int\frac{d^dl}{(2\pi)^d}\frac{d^dk}{(2\pi)^d}\frac{{\cal N}_{P}}{t\,l^2(l+p_3+p_4)^2(l+p_4)^2(k-l)^2k^2 (k-p_1-p_2)^2(k-p_1)^2} \label{DBplanar}\\
 \textup{\bf{DB}}_{NP}(s,t,u)&=\int\frac{d^dl}{(2\pi)^d}\frac{d^dk}{(2\pi)^d}\frac{{\cal N}_{NP}}{t\, l^2(l+p_3+p_4)^2(l+p_4)^2(k-l)^2k^2 (k-l+p_2)^2(k-p_1)^2} \label{DBnon-planar}\\
 \textup{\bf{DT}}_P(s)&=\int\frac{d^dl}{(2\pi)^d}\frac{d^dk}{(2\pi)^d}\frac{s^2}{l^2(l+p_3+p_4)^2(k-l)^2k^2(k-p_1-p_2)^2}\\
 \textup{\bf{DT}}_{NP}(s)& = \int\frac{d^dl}{(2\pi)^d}\frac{d^dk}{(2\pi)^d}\frac{s\, (l-p_2)^2}{l^2(l+p_3+p_4)^2(k-l)^2 k^2 (k-l+p_2)^2}
\end{align}
where the numerators are given by
\begin{align}
 {\cal N}_{P}&=\left[s\, \Tr(l\,p_1\,p_4)+l^2\,\Tr(p_1\,p_2\,p_4)\right]\left[s\,\Tr(k\,p_1\,p_4)+k^2\, \Tr(p_1\,p_2\,p_4)\right]\\
 {\cal N}_{NP}&=\left[s\, \Tr(l\,p_1\,p_4)+l^2\,\Tr(p_1\,p_2\,p_4)\right] \left[(l-p_2)^2\, \Tr(k\,p_1\,p_4)+ k^2\,\Tr(p_1\,(l-p_2)\,p_4)\right]
\end{align}

The first step towards the reduction of these integrals is to rewrite the complicated numerators in terms of a sum of inverse propagators. In order to do this let us introduce, following \cite{Smirnov:1999gc}, the most general double-box integral with seven propagators and two irreducible numerators. For the planar case
\begin{align}\label{generalP}
 &G_P(a_1,a_2,a_3,a_4,a_5,a_6,a_7,-a_8,-a_9)=\\ &\int\frac{d^dl}{(2\pi)^d}\frac{d^dk}{(2\pi)^d}\frac{[(l-p_1)^2]^{a_8}\, [(k+p_4)^2]^{a_9}}{[k^2]^{a_1}\, [(k-p_{12})^2]^{a_2}\,[l^2]^{a_3}\, [(l+p_{34})^2]^{a_4}\, [(l+p_4)^2]^{a_5}\, [(k-l)^2]^{a_6}\, [(k-p_1)^2]^{a_7}}\nonumber
\end{align}
In the non-planar case
\begin{align}\label{generalNP}
 &G_{NP}(a_1,a_2,a_3,a_4,a_5,a_6,a_7,-a_8,-a_9)=\\ &\int\frac{d^dl}{(2\pi)^d}\frac{d^dk}{(2\pi)^d}\frac{[(l-p_1)^2]^{a_8}\, [(k+p_3)^2]^{a_9}}{[k^2]^{a_1} [(k-p_{12})^2]^{a_2}[l^2]^{a_3} [(l-k-p_3)^2]^{a_4} [(l+p_4)^2]^{a_5} [(k-l)^2]^{a_6} [(k-p_1)^2]^{a_7}}\nonumber
\end{align}
In this notation the scalar double-box integral is represented as $G(1,1,1,1,1,1,1,0,0)$ and inverse propagators in the numerator lower one of the nine indices. We can then introduce the action of lowering operators such that
\begin{equation}
 G(1,1,1,1,1,1,1,0,0)=[\low{1}]^{a_1-1} G(a_1,1,1,1,1,1,1,0,0)
\end{equation}
For the planar double-box \eqref{DBplanar} the decomposition was first determined in \cite{CH} and, in our notation, it reads
\begin{equation}\label{decP}
 \textup{\bf{DB}}_P(s,t)=\frac12\, (s^2\, \low{9}\low{8}+s^2\, \low{7}\low{5}-s^2 t\, \low{6}+s\, t\, \low{4}\low{1}+s\, t\, \low{3}\low{2})\, G_P(1,1,1,1,1,1,1,0,0)
\end{equation}
The decomposition of \eqref{DBnon-planar} looks more complicated
\begin{equation} \label{decNP}
 \textup{\bf{DB}}_{NP}(s,t,u)=\frac12\, \low{O}\, G_{NP}(1,1,1,1,1,1,1,0,0)
\end{equation}
with
\begin{align}
\low{O}&=-s^2 t u\, \low{3} + 2\, s t u\, \low{1} \low{3} - u t\, [\low{1}]^2 \low{3} + 
  s t u\, \low{2} \low{3} + s t\, \low{1} \low{2} \low{3} -  t^2\, \low{1} \low{2} \low{3} \nonumber\\
  &+s^2 t\, \low{3} \low{7} -  s t\, \low{1} \low{3} \low{7} -2\, s t\, \low{2} \low{3} \low{7}
  + s t^2\, \low{1} \low{9} -   t^2\, [\low{1}]^2 \low{9} + s^2 t\, \low{3} \low{9}  \nonumber\\
  &-s t u\, \low{3} \low{9} -2\,  s t\, \low{1} \low{3} \low{9} -t^2\, \low{1} \low{3} \low{9} -  s t\, \low{2} \low{3} \low{9} -t^2\, \low{1} \low{4} \low{9} +  t^2\, \low{1} \low{5} \low{9}\nonumber \\
  &-  s t^2\, \low{6} \low{9} + 
  t^2\, \low{1} \low{6} \low{9} +  s t\, \low{3} \low{7} \low{9} +s t\, \low{5} \low{7} \low{9} -  s^2 t\, \low{8} \low{9}+ s t\, \low{1} \low{8} \low{9}\nonumber \\
  &+  s t\, \low{2} \low{8} \low{9} + 
  t^2\, \low{1} [\low{9}]^2 +  s t\, \low{3} [\low{9}]^2 - 
  s t\, \low{8} [\low{9}]^2
\end{align}

Once we obtain expressions like \eqref{decP} and \eqref{decNP} we can use the algorithm \texttt{FIRE} to reduce them to linear combinations of the following master integrals:
\begin{align}
\text{\bf SUNSET}(q^2)
&
= \begin{minipage}{45px}\sunset\end{minipage}
  =
  \frac{1}{(4\pi)^3}\left(\frac{-q^2 }{4\pi\mu^2}\right)^{-2\epsilon}
  \frac{\Gamma(\frac{1}{2}-\epsilon)^3\Gamma(2\epsilon)}
  {\Gamma(\frac{3}{2}-3\epsilon)}\,;\label{eq:sunset}\\
\text{\bf TRI}(q^2) &
=
\begin{minipage}{45px}\tri\end{minipage}
  =
  -\frac{(-q^2)^{-1}}{(4\pi)^3}
  \left(\frac{-q^2 }{4\pi\mu^2}\right)^{-2\epsilon}
  \frac{2\,\Gamma
    (
    \frac{1}{2}-\epsilon)^2\,\Gamma(-2\epsilon
    )\,
    \Gamma(\frac{3}{2}+\epsilon)\,
  \Gamma(2 \epsilon )}
  {(1 + 2 \epsilon) \Gamma
    ( \frac{1} {2} - 3 \epsilon )}\, ;
    \label{eq:tri}\\
\text{\bf GLASS}(q^2) &
=
\begin{minipage}{72px}\glass\end{minipage}
  =\frac{(-q^2)^{-1}}{(4\pi)^3}
  \left(\frac{-q^2 }{4\pi\mu^2}\right)^{-2\epsilon}
  \frac{\Gamma\left(\frac{1}{2}
  - \epsilon\right)^4\Gamma\left(\frac{1}{2}
  +
  \epsilon\right)^2}{\Gamma\left(1-2\epsilon\right)^2}\, ;
  \label{eq:glass}\\
 \text{\bf TrianX}(q^2)&\begin{aligned}[t] &
  =
  \begin{minipage}{72px}\centering\trianx\end{minipage}
  =\frac{(-q^2)^{-3}}{16\pi^2}
  \left(\frac{-q^2 e^{\gamma_E}}{4\pi\mu^2}\right)^{-2\epsilon}\,
  \biggl[ \frac {1} {\epsilon^2}
    + \frac{(3 + 8\log{2})}{4\epsilon}\\
  & -\left (\frac{27}{2} + \frac23 \pi^2
  + 4\log^2{2}-9\log{2}\right)
  + {\cal O}(\epsilon)\biggr]\, ;
\end{aligned}
\label{eq:trianx}\\
\text{\bf DIAG}(s,t)&\begin{aligned}[t] &
  =
  \begin{minipage}{50px}\centering\diag\end{minipage}
  =\frac{(-s)^{-2}}{16\pi^2}\left(\frac{-s e^{\gamma_E}}{4\pi\mu^2}\right)^{-2\epsilon} \biggl[ \frac {y} {2\,\epsilon}
    + y+ \frac12\, y\, \log y+ {\cal O}(\epsilon)\biggr] \, ;
\end{aligned}
\label{eq:diag}
\\
\text{\bf MUG}(s,t)&\begin{aligned}[t] &
  =
  \begin{minipage}{50px}\centering\mug\end{minipage}
  =\frac{(-s)^{-2}}{16\pi^2}\left(\frac{-s e^{\gamma_E}}{4\pi\mu^2}\right)^{-2\epsilon} \biggl[ \frac {y} {16\,\epsilon^2}
    +\frac{2\,y-\log2}{4\e}+ {\cal O}(\epsilon^0)\biggr]\text{\footnotemark};
\end{aligned}
\label{eq:mug}
\end{align}
\footnotetext{The finite part of this integral, which we computed to be a complicated combination of derivatives of hypergeometric functions, turned out to be irrelevant for our computation.}

\begin{align}
\text{\bf LAD}(s,t) &
\begin{aligned}[t] &
  = 
  \begin{minipage}{72px}\centering\ladder\end{minipage}
  =\frac{(-s)^{-4}}{16\pi^2}\left(\frac{-s e^{\gamma_E}}{4\pi\mu^2}\right)^{-2\epsilon} \biggl[ \frac {2\, y+3\, y^2} {2\,\epsilon^2} \\
  &+ \frac{27\,y+20\,y^2}{4\,\e}
    +\frac{(2\,y+3\,y^2)\,(2\, \log2+\log y)}{2\, \e} \\&
    - (2\,y+3y^2)\left(\frac{2\,\pi^2}{3}+ 2\,\log^2 2 \right) 
    + \frac{23}{2}y-7y^2 \\& 
    + (13\,y+20\,y^2)\log2 + 7\, (y+y^2)\, \log y + {\cal O}(\epsilon)\biggr]\, ;
\end{aligned}
\label{eq:ladder}
\\
\nonumber\\
\text{\bf LADn}(s,t)&\begin{aligned}[t] &
  =
  \begin{minipage}{72px}\centering\laddern\end{minipage}
  =-\frac{(-s)^{-3}}{16\pi^2}\left(\frac{-s e^{\gamma_E}}{4\pi\mu^2}\right)^{-2\epsilon} \biggl[ \frac {y} {2\,\epsilon^2}+\frac{5}{8\e}\\
  &+ \frac{y\,(2\log2+\log y)}{2\,\e}-y\,\left(\frac{2\,\pi^2}{3}+2\, \log^2 2\right)+\left(4\,y+\frac32\right)\log2\\
  &-6\, y+\frac54+ {\cal O}(\epsilon)\biggr]\, ;
\end{aligned}
\label{eq:laddern}\\
\nonumber\\
\text{\bf NPL2}(s,t,u)&\begin{aligned}[t] &
  =
  \begin{minipage}{72px}\centering\npladdertwo\end{minipage}
  =-\frac{(-s)^{-5}}{16\pi^2}\left(\frac{-s e^{\gamma_E}}{4\pi\mu^2}\right)^{-2\epsilon} \biggl[ -\frac {5\, y^2+7\, y^3+2\, \frac{y^3}{x}} {2\,\epsilon^2}\\
  &-\frac{(5\, y^2+7\, y^3+2\, \frac{y^3}{x})\,(2\, \log2+\log \frac{y}{x})}{2\, \e} +\frac{267\, xy+35\, x^2y^2+\frac{35}{2}x^3y^3}{24\,\e}\\
  &+ \frac{-281\,y^2 - 352\,y^3-89\,\frac{y^3}{x}+588\, xy^2-70\,xy^3+70\, x^2y^3}{24\,\e}\\
  &+\left(5\, y^2+7\, y^3+2\, \frac{y^3}{x}\right)\left(\frac{\pi^2}{3}+2\log^2 2 \right)-\frac{19}{12}\, y^2-\frac1{12}\, y^3+\frac{35}{8} \frac{y^3}{x}\\
  &-\left(\frac{79}2\, y^2+48\, y^3+16\, \frac{y^3}{x}\right)\log2-\frac{242\, y^2+274\,y^3+ 77 \,\frac{y^3}{x}}{12}\log \frac{y}{x}  \\
  &\frac{+35\, xy(-6\,y+2\, y^2-2\, xy^2-\frac{\,x^2y^2}{2}-x y-3)}{12}\, \log \frac{y}{x}-\frac{211}{6}\, x y^3\\
  &-\frac{11}{4}\, x^2y^3+\frac{285}{4}\, xy-\frac{29}{2}\, x^2y^2+\frac{37}{48}\, x^3y^3+ (x\leftrightarrow y)+ {\cal O}(\epsilon)\biggr]\, ;
\end{aligned}
\label{eq:npladdertwo}
\end{align}

\begin{align}
\text{\bf NPL}(s,t,u)&\begin{aligned}[t]&
  =
  \begin{minipage}{72px}\centering\npladder\end{minipage}
  =\frac{(-s)^{-4}}{16\pi^2}\left(\frac{-s e^{\gamma_E}}{4\pi\mu^2}\right)^{-2\epsilon} \biggl[ -\frac {2\, y+5\, y^2+2\, \frac{y^2}{x}} {4\,\epsilon^2}\\
  &-\frac{(2\, y+5\, y^2+2\, \frac{y^2}{x})\,(2\, \log2+\log \frac{y}{x})}{4\, \e} + \frac{-25\,y-37\,y^2-12\,\frac{y^2}{x}}{8\,\e}\\
  &+\frac{35\,xy+\frac{5}{2}\, x^2 y^2}{8\, \e}+\left(2\, y+5\, y^2+2\, \frac{y^2}{x}\right)\left(\frac{\pi^2}{6}+\log^2 2 \right)\\
  &-\left(6\,y+\frac{31}{4}\,y^2+3\, \frac{y^2}{x}-\frac{15}{4}\, xy-\frac58\, x^2y^2\right)\log \frac{y}{x}  \\
  &-\left(\frac{23}{2}\,y+18\, y^2+8\,\frac{y^2}{x} \right)\log2+\frac{y}{2}+\frac{15}{4}\,y^2+\frac{47}{12}\frac{y^2}{x}\\
  &-\frac{15}{4}\, xy^2+\frac{33}{2}\, xy-\frac1{24} x^2y^2+ (x\leftrightarrow y)+ {\cal O}(\epsilon)\biggr]
\end{aligned}
\label{eq:npladder}
\end{align}
In this list we introduced the variables $y=\frac{s}{t}$ and $x=\frac{s}{u}$. The integral $\text{\bf LADn}(s,t)$ is a double-box with an irreducible numerator given by an inverse propagator with momentum equal to the sum of the two momenta running in the direction of the arrows. In the notation of \eqref{generalP} it is given explicitly by
\begin{equation}
 \text{\bf LADn}(s,t)=G_P(1,1,1,1,1,1,1,-1,0)
\end{equation}
For the non-planar case, instead of taking a master integral with an irreducible numerator we chose to use $\text{\bf NPL2}(s,t,u)$, whose computation turned out to be easier. $\text{\bf NPL2}(s,t,u)$ is given explicitly by
\begin{equation}
 \text{\bf NPL2}(s,t,u)=G_{NP}(1,1,1,1,1,2,1,0,0)
\end{equation}
The non-planar integral with one irreducible numerator can be computed using\\ $\text{\bf NPL2}(s,t,u)$ as master integral and the explicit relation is\footnote{All the expressions here are expanded to the order in $\e$ necessary to get the right finite part.}
\begin{align}
 &\text{\bf NPLn}(s,t,u)=G_{NP}(1,1,1,1,1,1,1,-1,0)\nonumber \\
 &=\frac{1}{t-u}\Big[-\frac{stu}{72}(9-15\, \e+23\, \e^2)\,\text{\bf NPL2}(s,t,u)-\Big(tu-\frac{s^2}{108}(45-6\e+8\e^2)\Big)\,\text{\bf NPL}(s,t,u)\nonumber \\
 &-\frac{u}{24\,st}(9\,t\,(7+5\,\e)+s\,(39+29\e))\,\text{\bf DIAG}(s,t)-\frac{t}{24\,su}(72\,s+t\,(129-13\, \e))\,\text{\bf DIAG}(s,u)\nonumber \\
 &+\frac{s}{3\,tu}(s\,(1+2\e)-t\, (3+2\e))\,\text{\bf DIAG}(t,u)-\Big(t+\frac{s}{8}(5+\e-\e^2)\Big)\, \text{\bf TrianX}(s) \nonumber \\
 &-\e\,\frac{3\,s\,(9-11\,\e)+t\,(39-49\,\e)}{6t}\text{\bf MUG}(s,t)-\e\,\frac{s\,(60-128\,\e)+t\,(105-239\,\e)}{6u}\text{\bf MUG}(s,u)\nonumber \\
 &-\e\,\frac{16\, st\,\e-8\, t^2(3+4\,\e)+s^2(9+25\, \e)}{4\,stu}\text{\bf TRI}(s)-\e\,\frac{26\, s^2+23\, st-41\, t^2}{8\,s^2tu}\text{\bf SUNSET}(s)\nonumber\\
 &-\e\,\frac{129\, s^2+191\, st+126\, t^2}{24\,st^2u}\text{\bf SUNSET}(t)-\e\,\frac{-16\, s^2+227\, st+258\, t^2}{24\,stu^2}\text{\bf SUNSET}(u)\Big]
\end{align}
The decomposition of the planar integral \eqref{DBplanar} is simply given by
\begin{align}
\textup{\bf{DB}}_{P}(s,t)&=\frac{s^3 t}{4}\, \text{\bf LAD}(s,t)+\frac{3\, s^3}{4}\, \text{\bf LADn}(s,t)-\frac{7\, s\, (s+t)}{2}\,\text{\bf DIAG}(s,t)\nonumber\\
 &-2\,(1-5\,\e)\, \text{\bf SUNSET}(s)+8\,\e\,(1-2\e)\, s^2\,  \text{\bf MUG}(s,t)-\e\,\frac{18\, s}{t}\,\text{\bf SUNSET}(t)\nonumber \\
 &-17\, s\,\e\,(1+2\e)\, \text{\bf TRI}(s)
\end{align}
The explicit expression for the decomposition of the non-planar integral \eqref{DBnon-planar} reads 
\begin{align}
 &\textup{\bf{DB}}_{NP}(s,t,u)
 =(8\e^2-2\e+1)\, \frac{s}{4}\,\Big[-(2s^2+(t-u)^2)\,\text{\bf NPLn}(s,t,u)\nonumber\\
 &+  t\, u\, (t-u)\,\text{\bf NPL}(s,t,u)-t\,(2u-s)\,\text{\bf TrianX}(s)\Big]\nonumber \\
 &+3\, s\,\e\,(1-4\e)\left(\frac{3\,s^2}{u}-4u\right)\text{\bf MUG}(s,t)-\frac{u}{12}(15\,s-42\,t+(-28\,s+80\, t )\,\e)\,\text{\bf DIAG}(s,t)\nonumber\\
 &+\frac{1}{12\, u} \left(48\,t^3(1-4\e)+24\, st^2(5-18\e)+s^3(9-20\e)+2\, s^2t\,(45-136\e)\right) \text{\bf DIAG}(t,u) \nonumber \\
 &+\frac{1}{12\, u} \left(48\,s^3(1-4\e)-st^2(3+92\e)+2\, t^3(3-4\e)+3\, s^2t\,(25-114\e)\right)\text{\bf DIAG}(s,u)\nonumber \\
 &-\frac{s}{2\,u}\, \e\, (15\, s+6\, t+64\, u\,\e)\text{\bf TRI}(s)-4\,\Big[\text{\bf SUNSET}(s)+\text{\bf SUNSET}(t)+\text{\bf SUNSET}(u)\Big]\nonumber \\
 & +\e\, \left(23+\frac{6\,t}{u}-\frac{18\, t}{s}\right)\text{\bf SUNSET}(s)+\e\,\left(\frac{41}{2}+\frac{3\, t}{u}-\frac{6\, u}{t}\right)\text{\bf SUNSET}(t) \nonumber \\
 &+\e\, \left(40+\frac{9\,t}{2\, u}-\frac{9\, t^2}{2\, u^2}\right)\text{\bf SUNSET}(u)
\end{align}
Using the results of the master integrals listed in \eqref{eq:sunset}-\eqref{eq:npladdertwo} we obtain, for the planar integrals,
\begin{align}
\textup{\bf{DB}}_P(s,t)&= -\frac{1}{16\pi^2}\left(\frac{-s e^{\g_E}}{4\pi\mu^2}\right)^{-2\e}\left(\frac{1}{4\e^2}-\frac{1-\log2-\tfrac12 \log\frac{s}{t}}{2\e} - 1 - 2\z_2 - \log^2 2+\mathcal{O}(\e)\right)\, \label{db}\\
\textup{\bf{DT}}_P(s)&=\frac{1}{16\pi^2}\left(\frac{-s e^{\g_E}}{4\pi\mu^2}\right)^{-2\e}\left(-\frac{1}{2\e}-1+\mathcal{O}(\e)\right)\label{dt}
\end{align}
and, for the non-planar case
\begin{align}
\textup{\bf{DB}}_{NP}(s,t,u)& = -\frac{1}{16\pi^2}\left(\frac{-s e^{\g_E}}{4\pi\mu^2}\right)^{-2\e}\left(\frac{1}{8\e^2}+\frac{2\log2+\log\frac{u}{t}}{8\e} + \frac12\log\frac{s}{u}
\right. \nonumber\\& \left. ~~~ 
-\frac12\z_2 -\frac12\log^2 2 -1+\log2 + {\cal O}(\epsilon)\right)\, \label{dbnp}\\
 \textup{\bf{DT}}_{NP}(s)& = -\frac{1}{16\pi^2}(1-\log2)  + {\cal O}(\epsilon)\, \label{dtnp} \\
 \textup{\bf{XT}}(s)& = -\frac{1}{16\pi^2}\left(\frac{-s e^{\g_E}}{4\pi\mu^2}\right)^{-2\e}\left(\frac{1}{4\e^2}+\frac{2\log2}{4\e}-\z_2-\log^2 2  + {\cal O}(\epsilon) \right)
\end{align}
The integral $\textup{\bf{XT}}(s)$ was defined in \eqref{XT} and details about its reduction to master integrals can be found in \cite{Brandhuber:2013gda}. Here we quote only the final result.

\section{Three-dimensional master integrals for four-point massless\\ scattering}\label{app:3}

In this appendix we give some details for the explicit evaluation of the master integrals involved in the computation of the two-loop four-point amplitude and Sudakov form factor.
Among the results listed in appendix \ref{app:2} the expressions \eqref{eq:sunset}-\eqref{eq:trianx} were already given in \cite{Brandhuber:2013gda}. Therefore we will focus on the calculation of the master integrals \eqref{eq:diag}-\eqref{eq:npladdertwo}.

\begin{figure}[htbp]
\begin{center}
\includegraphics[width=13 cm]{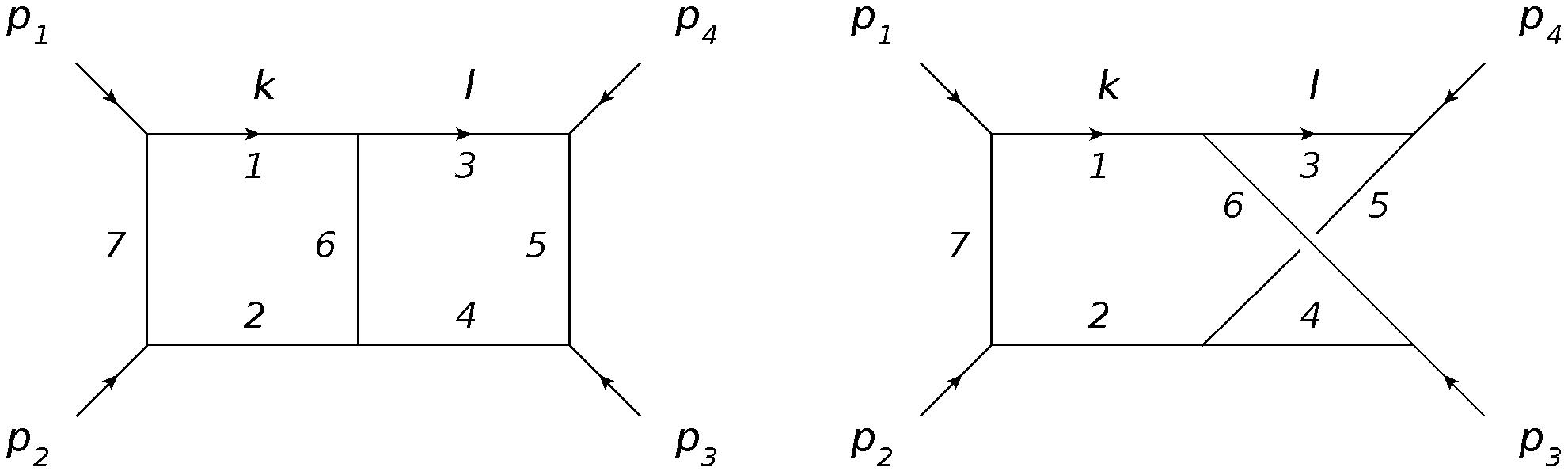}
\caption{Planar and non-planar double-box. The numbers indicate the index $a_i$ associated to any propagator in \eqref{generalP} and \eqref{generalNP}.}
\label{doubleboxes}
\end{center}
\end{figure}

The building blocks are the two double-box integrals introduced in \eqref{generalP} and \eqref{generalNP} and represented in Figure \ref{doubleboxes}. All our master integrals can be understood as special cases of those two general expressions. In fact we don't even need the be as general as in \eqref{generalP} and \eqref{generalNP}. Indeed we may restrict ourselves to values of the indices corresponding to the diagrams \eqref{eq:ladder}, \eqref{eq:laddern}, \eqref{eq:npladder} and \eqref{eq:npladdertwo}. We compute the integrals starting from their Mellin-Barnes representation. 

For the planar double-box a general representation of $G_P(a_1,a_2,a_3,a_4,a_5,a_6,a_7,a_8,0)$ is given in \cite{Smirnov:2004ym}. In $d=3-2\e$ and with the leg labelled as in Figure \ref{doubleboxes} it reads
\begin{equation}
 G_P(a_1,a_2,a_3,a_4,a_5,a_6,a_7,a_8,0)=-\frac{(-1)^a}{(4\pi)^d\, (-s)^{a-3+2\e}} F(y,\e)
\end{equation}
with $y=\frac{s}{t}$ and 
\begin{align}
 F(y,\e)&=\frac{1}{\prod_{l=3}^7 \Gamma(a_l)\, \Gamma(3-a_{3456}-2\e)}\int_{-i\infty}^{+i\infty} \prod_{j=1}^4\left(\frac{dz_j}{2\pi i}\right)\, y^{-z_1} \frac{\G(-z_1-z_2-z_3-z_4)}{\G(a_8-z_1-z_2-z_3-z_4)} \nonumber \\
 &\times \frac{\G(-z_1)\, \G(a_7+z_1)\, \G(z_2+z_4)\, \G(z_3+z_4)\, \G(a_{1278}-\frac32+\e+z_4)\, \G(a_6+z_1-z_4)}{\G(a_1+z_3+z_4)\, \G(a_2+z_2+z_4)\,\G(3-a_{1278}-2\e+z_1-z_4)}\nonumber \\
 &\times\G(a_5+z_1+z_2+z_3+z_4)\, \G(a_8-z_2-z_3-z_4)\, \G\left(a_{3456}-\frac32+\e+z_1-z_4\right)\nonumber \\ 
 &\times\G\left(\frac32-a_{356}-\e-z_1-z_2\right)\, \G\left(\frac32-a_{456}-\e-z_1-z_3\right)\nonumber \\
 &\times\G\left(\frac32-a_{178}-\e+z_2\right)\, \G\left(\frac32-a_{278}-\e+z_3\right) \label{generalMBP}
\end{align}

This Mellin-Barnes representation can be used to evaluate most of the master integrals listed in appendix \ref{app:2}. In particular the integrals \eqref{eq:diag} and \eqref{eq:mug} can be computed by carefully taking the limit of some indices to zero. Integral \eqref{eq:ladder} was first computed, starting from this Mellin-Barnes representation, in \cite{Bianchi:2012ez}. Adding an irreducible numerator, i.e. setting $a_8=-1,\ a_9=0$ and all the other indices to one in \eqref{generalMBP}, gives the master integral \eqref{eq:laddern}. We computed this integral with the help of the routine \texttt{MB.m} \cite{Czakon:2005rk} which reduced the four-fold Mellin-Barnes representation to a one-fold integral. The latter was solved using Barnes lemmas and their corollaries.

Moving to the non-planar case, as we mentioned in appendix \ref{app:2} the computation of the non-planar double-box $\text{\bf NPL2}(s,t,u)$ (which can be obtained from \eqref{generalNP}, setting the first six indices to 1 and $a_7=2$) turns out to be simpler than the one with an irreducible numerator. Indeed we can introduce a generalization for arbitrary values of the indices of the Mellin-Barnes representation given in \cite{Tausk:1999vh}.
Explicitly
\begin{equation}
 G_{NP}(a_1,a_2,a_3,a_4,a_5,a_6,a_7,0,0)=-\frac{(-1)^a}{(4\pi)^d\, (-s)^{a-3+2\e}} f(x,y,\e)
\end{equation}
with $y=\frac{s}{t}$, $x=\frac{s}{u}$ and
\begin{align}\label{generalMBNP}
f( x , y, \e ) &=\frac{\G(\frac32-\e-a_{45}) \G(\frac32-\e-a_{67})}{\G(\frac92-3\e-a)\G(3-2\e-a_{4567})}
\int_{-i\infty}^{+i\infty}\, \prod_{j=1}^4\, \left(\frac{dz_j}{2\pi i}\right)\,
x^{-z_1} y^{-z_2} \,\Gamma (z_1+z_2+a_2)
\nonumber\\
&\times  \Gamma (-z_1)\, \Gamma (-z_2)\, \Gamma (-z_3)\,\Gamma (-z_4)\,\Gamma \left(-3+2 \epsilon +z_1+z_2+z_3+z_4+a\right)
\nonumber \\
&\times\Gamma (z_1+z_3+a_5)\, \Gamma (z_2+z_3+a_7)\,  \Gamma (z_1+z_4+a_6)\, \Gamma (z_2+z_4+a_4)
\nonumber\\
&\times\Gamma \left(3-2 \epsilon -z_1-z_2-z_3-a+a_3\right)\,
\Gamma \left(3-2 \epsilon -z_1-z_2-z_4-a+a_1\right)  
\nonumber\\&\times
 \frac{\Gamma \left(-\frac{3}{2}+\epsilon +z_1+z_2+z_3+z_4+a_{4567}\right)}{
\Gamma (z_1+z_2+z_3+z_4+a_{67})\, \Gamma(z_1+z_2+z_3+z_4+a_{45})}
\end{align}

This Mellin-Barnes representation can be used to compute the integrals $\text{\bf NPL}(s,t,u)$ and  $\text{\bf NPL2}(s,t,u)$. In both cases, thanks to the $\G(3-2\e-a_{4567})$ in the denominator, the four-fold integral can be reduced by deforming the integration contour and analytically continuing to the region close to $\e=0$. This task can be performed automatically with the routine \texttt{MB.m} which returns a complicated combination of one-fold Mellin-Barnes integrals, whose evaluation, though long and tedious, requires just the application of Barnes lemmas and their corollaries. The results are reported in \eqref{eq:npladder} and \eqref{eq:npladdertwo}.

\section{One-loop $d$-dimensional cuts}\label{app:dcut}
In this appendix we prove that the combination $I(1,2,4,3)-I(1,4,2,3)$ has a non-vanishing $d$-dimensional cut in the $u$-channel and by explicit integration we verify that at first order in the parameter $\e$ the discontinuity reproduces the imaginary part in \eqref{ident}. In order to achieve this result we generalize to the three-dimensional case the procedure outlined in \cite{Anastasiou:2006gt, Anastasiou:2006jv}. Let us start with a generic one-loop integrand with $n$ propagators and a cut in the $K^2$ channel
\begin{equation}
 \Delta {\cal I}^{(a)}_n  = \int d^{3-2\e}\, \hat \ell\, {
(2\hat\ell\cdot T)^a\, \delta( \hat\ell^2)\, \delta((\hat \ell-K)^2)\over
\prod_{i=1}^{n-2} (\hat
\ell-K_i)^2}~.~~~\label{1loop-gen-integrand-cut}
\end{equation}
where $K_i$ and $T$ are generic combination of the external legs and therefore inherently three-dimensional objects. On the contrary, $\hat{l}$ is a $3-2\e$-dimensional vector which can be decomposed as $\hat \ell=\tilde \ell+\mu$, where $\tilde\ell$ is the pure three-dimensional part while
$\mu$ is the $(-2\e)$-dimensional part. This gives
\begin{equation}
\Delta {\cal I}^{(a)}_n=\int d^{-2\epsilon}\mu\int d^3\tilde \ell\,   {(2\tilde\ell\cdot T)^a\,\delta(
\tilde\ell^2-\mu^2) \delta((\tilde \ell-K)^2-\mu^2)\over\prod_{i=1}^{n-2} ((\tilde
\ell-K_i)^2-\mu^2)}~.~~~
\end{equation}
We further rewrite $\tilde\ell$ and $\tilde\ell=\ell+z\, K$ with $\ell^2=0$ such that
\begin{equation}
\int d^3\tilde \ell\,  \delta(\tilde\ell^2-\mu^2) \delta((\tilde \ell-K)^2-\mu^2)=\int dz
d^3\ell\, \delta(\ell^2)\, 2\ell\cdot
K\, \delta( z^2K^2+2z\ell\cdot
K-\mu^2)\, \delta((1-2z)K^2-2\ell\cdot
K).
\end{equation}
So far the procedure is exactly identical to the four-dimensional case, however here comes the major difference. The integral over the loop momentum, once the on-shell delta function is enforced, contains only two residual degrees of freedom, one less than the corresponding four-dimensional case. This implies that a different spinor integration has to be introduced. This can be done writing $\ell^{\a\b}=t \l^\a \l^\b$ and fixing the normalization to correct the mismatch mentioned in section 2.1 of \cite{Anastasiou:2006gt}.  This leads to
\begin{equation}
\int d^3\ell\, \delta(\ell^2)=\frac{1}{8 \p} \int_0^\infty dt \int \braket{\l\,d\l}  
\end{equation}
We notice that, under rescaling of $\l$, we need to ask the factor $t$ to transform with weight -2 in order to leave the loop momentum invariant. This immediately implies that the measure is invariant under rescaling of $\l$ as required. This is also the reason for the difference in the power of $t$ compared to the four-dimensional case. In this case the integration contour is simply the real axis.

Solving the $\d$-functions to eliminate the integrals over $t$ and $z$ we obtain the following final formula
\begin{equation}
\Delta {\cal I}^{(a)}_{n}= \frac{1}{8\p} \int d^{-2\e} \mu \int
\braket{\l\,d\l} \frac{(-)^{n-2}\, [(1-2z)\, K^2]^{a-n+2}\, 
\braket{\l|R|\l}^a}{
\braket{\l|K|\l}^{a-n+3}\, \prod_{i=1}^{n-2}\braket{\l|Q_i|\l}}~,~~~
\label{1loop-uni-gen} 
\end{equation}
with the following identifications 
\begin{align}
z&={1-\sqrt{1-y}\over 2}~,& y&\equiv {4\mu^2\over K^2}~,\\
 R &\equiv T+ {z (2K\cdot T)\over (1-2z) K^2} K~~,& Q_i&\equiv
K_i+{z (2K\cdot K_i)-K_i^2\over (1-2z) K^2} K
\end{align}
In the following we are going to be interested only in some very simple cases, i.e. the one-mass scalar triangle and the massless scalar box. We will see how those results turn out to be particularly simple and reproduce the all-order in $\e$ imaginary part of \eqref{eq:triangle} and \eqref{eq:box}.
\subsection{Triangle}
In the case of the scalar triangle formula \eqref{1loop-uni-gen}, setting $a=0$ and $n=3$, reduces to
\begin{equation}
 \mathbf{T}(K^2)|_{K^2\textup{-cut}}=\frac{1}{8\p}\int d^{-2\e}\mu \int
\braket{\l\,d\l} \frac{-1}{\braket{\l|Q_1|\l}\, (1-2z)\, K^2}
\end{equation}
which can be further simplified enforcing the condition $(K-K_1)^2=0$ 
\begin{equation}
 \mathbf{T}(K^2)|_{K^2\textup{-cut}}=\frac{1}{8\p}\int d^{-2\e}\mu \int
\braket{\l\,d\l} \frac{-1}{\braket{\l|(1-z)K_1+z K_2|\l} K^2}
\end{equation}
with $K_1+K_2=K$. The spinor integration simply gives
\begin{equation}
 \mathbf{T}(K^2)|_{K^2\textup{-cut}}=\frac{-1}{8\, (K^2)^{\frac32}}\int d^{-2\e}\mu \frac{1}{\sqrt{z(1-z)}}
\end{equation}
Using
\begin{equation}
 d^{-2\e}\mu=\frac{(4\p)^\e}{\Gamma(-\e)} \left(\frac{K^2}{4}\right)^{-\e} y^{-1-\e} dy
\end{equation}
we get 
\begin{equation}
 \mathbf{T}(K^2)|_{K^2\textup{-cut}}=\frac{-(4\p)^\e}{\Gamma(-\e)\, 4\, (K^2)^{\frac32}} \left(\frac{K^2}{4}\right)^{-\e} \int_0^1 d y\,  y^{-\frac32-\e}=\frac{(4\p)^\e}{2\, (1+2\e)\,\Gamma(-\e)\, (K^2)^{\frac32}} \left(\frac{K^2}{4}\right)^{-\e}
\end{equation}
We notice that the result is consistently of order $\e$ due to the factor $\frac{1}{\Gamma(-\e)}$ and that it can be obtained starting from equation \ref{eq:triangle} using the identity
\begin{equation}\label{eq:im}
 \textup{Im}[(-s)^{-\frac32-\e}]=s^{-\frac32-\e} \cos \p \e 
\end{equation}
valid in the cut kinematics $s>0$.
\subsection{Massless box}
We will consider the box with ordered external momenta $K_i$ for $i=1,...,4$ and $K_i^2=0$. This admit cuts in the $s$-channel $K=K_1+K_2$ and in the t-channel $K=K_1+K_4$. However the scalar box is surely symmetric in $s$ and $t$ and without loss of generality we can consider just the $s$-channel cut. Therefore formula \eqref{1loop-uni-gen}, setting $a=0$ and $n=4$, reduces to  
\begin{equation}
 \mathbf{B}(s,t) |_{s\textup{-cut}}=\frac{1}{8\p}\int d^{-2\e}\mu \int
\braket{\l\,d\l} \frac{\braket{\l|K|\l}}{\braket{\l|Q_1|\l}\, \braket{\l|Q_2|\l}\, ((1-2z)\, s)^2}
\end{equation}
with $Q_1=K_1+\frac{z (2K\cdot K_1)}{(1-2z) s} K$ and $Q_2=-K_4-\frac{z (2K\cdot K_4)}{(1-2z) s} K$. Using the identities $2K\cdot K_1=s$ and $2K\cdot K_4=-s$ the integrand simplifies to
\begin{equation}
 \mathbf{B}(s,t) |_{s\textup{-cut}}=\frac{1}{8\p\, s^2 }\int d^{-2\e}\mu \int
\braket{\l\,d\l} \frac{\braket{\l|K|\l}}{\braket{\l|\tilde Q_1|\l}\, \braket{\l|\tilde Q_2|\l}}
\end{equation}
with $\tilde Q_1=(1-z)K_1+z\,K_2$ and $\tilde Q_2=-(1-z)K_4-z\,K_3$. The spinor integration yields
\begin{equation}
 \mathbf{B}(s,t) |_{s\textup{-cut}}=\frac{1}{8\, s^2 }\int d^{-2\e}\mu \frac{1}{\tilde Q_1\cdot\tilde Q_2+|\tilde Q_1||\tilde Q_2|}\, \left(\frac{K\cdot\tilde Q_1}{|Q_1|}+\frac{K\cdot\tilde Q_2}{|Q_2|}\right)
\end{equation}
Moreover using $K\cdot\tilde Q_2=K\cdot\tilde Q_1=s$ and $|\tilde Q_1|=|\tilde Q_2|=\sqrt{z(1-z)s}$ we obtain
\begin{equation}
 \mathbf{B}(s,t) |_{s\textup{-cut}}=\frac{1}{4\, s^{\frac32} }\int d^{-2\e}\mu \frac{1}{4\, s\, z(1-z)-t\,(1-2z)^2 }\frac{1}{\sqrt{z(1-z)}}
\end{equation}
and switching to $y$ variables 
\begin{equation}
 \mathbf{B}(s,t) |_{s\textup{-cut}}=\frac{(4\p)^\e}{\Gamma(-\e)\,2\, s^{\frac52}} \left(\frac{s}{4}\right)^{-\e} \int dy\, \frac{1}{y- x (1-y) }y^{-\frac32-\e}
\end{equation}
where $x=\frac{t}{s}$. The integration is still simple enough to be performed at all-order in $\e$, giving
\begin{equation}
 \mathbf{B}(s,t) |_{s\textup{-cut}}=\frac{(4\p)^\e}{(1+2\e)\, \Gamma(-\e)\, s^{\frac32}\, t} \left(\frac{s}{4}\right)^{-\e}\, _2F_1 \left(\begin{array}{c} 1, -1/2-\epsilon;\\ 1/2-\epsilon\end{array} \bigg| 1+\frac{s}{t} \right)
\end{equation}
Once again we notice that this cut could be obtained simply from equation \eqref{eq:box} using \eqref{eq:im}. Indeed in a regime with $s>0$ and $t<0$ the hypergeometric function is purely real and the only imaginary part comes from the pre-factor $(-s)^{-\frac32-\e}$. Let us comment on the fact that this identification of the expression of the cut with the imaginary part of the result of the integral is simple just at one loop where the identification does not involve a sum over the cuts. In general, given the result of an integral is not straightforward to find the expressions of the different cuts.

\subsection{u-channel cut of $I(1,2,4,3)-I(1,4,2,3)$}
Equipped with the cuts of the triangle and of the scalar box we can easily compute the $u$-channel cut of $I(1,2,4,3)-I(1,4,2,3)$ using the decomposition \eqref{decomposition}. The only subtlety comes in the mutual sign of the two decompositions. To understand how this works let us introduce the notation $I(s,t)$ for $I(1,2,3,4)$ given by the decomposition \eqref{decomposition}. Permuting legs 2 and 4 obviously implies exchanging $s$ with $t$, but in order to respect the symmetry property $I(1,2,3,4)=-I(1,4,3,2)$ one has to require $I(1,4,3,2)=-I(t,s)=-I(s,t)$. In a similar way we have $I(1,3,4,2)=I(s,u)=-I(1,2,4,3)$ and $I(1,4,2,3)=I(u,t)=-I(1,3,2,4)$. The result of this analysis is that
\begin{align}
 I(1,2,4,3)&-I(1,4,2,3)=-I(s,u)-I(u,t)\nonumber\\
 &=\frac{1}{2\, i \sqrt{s\,t\,u}}(2\, s^2u\, \TRIANGLE(s)+2\, t^2u\, \TRIANGLE(t)-2\, u^3\, \TRIANGLE(u)-u^2s^2\,\BOX(s,u)-u^2t^2\,\BOX(u,t) )
\end{align}
It is then clear that the $d$-dimensional $u$-channel cut of this expression is given by
\begin{align}
 I(1,2,4,3)-I(1,4,2,3)|_{u\textup{-cut}}&=\frac{u^2}{2\, i \sqrt{s\,t\,u}}\left(2\,u\,\TRIANGLE(u)|_{u\textup{-cut}}-s^2\,\BOX(s,u)|_{u\textup{-cut}}-t^2\,\BOX(u,t)|_{u\textup{-cut}}\right)\nonumber \\
 &=\frac{1}{2\, i \sqrt{s\,t}}\frac{(4\p)^\e}{(1+2\e)\, \Gamma(-\e)} \left(\frac{u}{4}\right)^{-\e}\left(u+s\, F(s)+t\, F(t)\right)
\end{align}
where $F(x)= {}_2F_1 \left(\begin{array}{c} 1, -1/2-\epsilon;\\ 1/2-\epsilon\end{array} \bigg| 1+\frac{u}{x} \right)$. Expanding in $\e$ the first order gives
\begin{align}
 I(1,2,4,3)-I(1,4,2,3)|_{u\textup{-cut}}&=\e \frac{\p}{4} +\mathcal{O}(\e^2)
\end{align}
in perfect agreement with \eqref{ident}. This proves that the rule of replacing $I(1,2,3,4)$ with \\$I(1,2,4,3)-I(1,4,2,3)$ does not work in the framework of $d$-dimensional cuts. This implies that the all-order in $\e$ expression for the one-loop amplitude is \eqref{eq:myamp}.

\vfill
\newpage

\bibliographystyle{nb}

\bibliography{biblio}

\end{document}